\documentclass[preprint,prc,letterpaper,11pt,twoside,tightenlines,nofootinbib,showpacs]{revtex4-1}
\usepackage[bookmarks,pdfhighlight=/O,colorlinks=false,pdfstartview=FitH]{hyperref}
\usepackage{amsmath,amssymb,mathrsfs,slashed,bm}
\usepackage{color}
\usepackage{graphicx}
\usepackage{epstopdf}

\newcommand{\ii}{\mathrm{i}}
\newcommand{\dd}{\mathrm{d}}
\newcommand{\MeV}{\mathrm{MeV}}
 \newcommand{\GeV}{\mathrm{GeV}}
\newcommand{\TeV}{\mathrm{TeV}}
\newcommand{\fm}{\mathrm{fm}}
\newcommand{\bvec}[1]{\ensuremath{\boldsymbol{#1}}}
\newcommand{\erw}[1]{\ensuremath { %
    \left \langle {#1} \right \rangle}}

\newcommand{\Lag}{\ensuremath{\mathscr{L}}}
\newcommand{\fslash}{\slashed}

\newcommand{\tildeint}[1]{\ensuremath{\int_{\R^3} \frac{\mathrm{d}^{3} #1}{2
      E(#1) \, (2\pi)^{3}}}}
\newcommand{\R}{\ensuremath{\mathbb{R}}}

\begin{document}

\preprint{}

\title{Heavy quark transport in heavy ion collisions at RHIC and LHC
  within the UrQMD transport model}

\author{Thomas Lang$^{1,2}$} \author{Hendrik van Hees$^{1,2}$} \author{Jan Steinheimer$^{3}$}
\author{Marcus Bleicher$^{1,2}$}
\affiliation{
  $^{1}\,$Frankfurt Institute for Advanced Studies
  (FIAS),Ruth-Moufang-Str. 1, 60438 Frankfurt am Main, Germany }
\affiliation{
  $^{2}\,$Institut f\"ur Theoretische Physik, Johann Wolfgang
  Goethe-Universit\"at, Max-von-Laue-Str. 1, 60438 Frankfurt am Main,
  Germany }
\affiliation{
  $^{3}\,$Lawrence Berkeley National Laboratory, 1 Cyclotron Road, Berkeley, CA 94720, USA
}

\date{\today}

\begin{abstract}
  We have implemented a Langevin approach for the transport of heavy
  quarks in the UrQMD hybrid model. The UrQMD hybrid approach provides
  a realistic description of the background medium for the evolution
  of relativistic heavy ion collisions. We have used two different sets
  of drag and diffusion coefficients, one based on a $T$-Matrix approach
  and one based on a resonance model for the elastic scattering of heavy
  quarks within the medium. In case of the resonance model we have
  investigated the effects of different decoupling temperatures of the
  heavy quarks from the medium, ranging between $130\;\MeV$ and
  $180\;\MeV$. We present calculations of the nuclear modification
  factor $R_{AA}$, as well as of the elliptic flow $v_2$ in Au+Au
  collisions at $\sqrt{s_{NN}}=200\; \text{GeV}$ and Pb+Pb collisions at
  $\sqrt{s_{NN}}=2.76 \; \TeV$. To make our results comparable to
  experimental data at RHIC and LHC we have implemented a Peterson
  fragmentation and a quark coalescence approach followed by the
  semileptonic decay of the D- and B-mesons to electrons. We find that
  our results strongly depend on the decoupling temperature and the
  hadronization mechanism.  At a decoupling temperature of $130\;\MeV$
  we reach a good agreement with the measurements at both, RHIC and LHC
  energies, simultaneously for the elliptic flow $v_2$ and the nuclear
  modification factor $R_{AA}$.
\end{abstract}

\maketitle


\section{Introduction}

\markright{Thomas Lang, Hendrik van Hees, Jan Steinheimer, and Marcus
  Bleicher. Heavy quark transport in heavy ion collisions}

One major goal of ultra-high-energy heavy-ion physics is to recreate the
phase of deconfined quarks and gluons (the Quark Gluon Plasma, QGP) as
it might have existed a few microseconds after the Big Bang. Various
experimental facilities have been built to explore the properties of
this QGP experimentally, while on the theory side a multitude of
(potential) signatures and properties of the QGP have been predicted 
\cite{Adams:2005dq,Adcox:2004mh,Muller:2012zq}.

Heavy quarks are an ideal probe for the QGP. They are produced in the
beginning of the collision in hard processes and therefore probe the
created medium during its entire evolution. When the system cools down
they hadronize, and their decay products can finally be detected. By
investigating heavy-quark observables we can thus explore the
interaction processes within the hot and dense medium. Two of the most
interesting observables are the nuclear modification factor, $R_{AA}$,
and the elliptic flow, $v_2$. Experimentally, the nuclear modification
factor shows a large suppression of the open heavy-flavor particles'
spectra at high transverse momenta ($p_T$) compared to the findings in
pp collisions. This indicates a high degree of thermalization also of
the heavy quarks with the bulk medium consisting of light quarks and
gluons and, perhaps at the later stages of the fireball evolution, the
hot and dense hadron gas. The measured large elliptic flow, $v_2$, of
open heavy-flavor mesons and the non-photonic single electrons or muons
from their semileptonic decay underlines this interpretation because it
indicates that heavy quarks take part in the collective motion of the
bulk medium. A quantitative analysis of the degree of thermalization of
heavy-quark degrees of freedom in terms of the microscopic
scattering processes may lead to an understanding of the 
mechanisms underlying the large coupling strength of the QGP and the
corresponding transport properties.

In this paper we explore the medium modification of heavy-flavor $p_T$
spectra, using a hybrid model, consisting of the Ultra-relativistic
Quantum Molecular Dynamics (UrQMD) model
\cite{Bass:1998ca,Bleicher:1999xi} and a full (3+1)-dimensional ideal
hydrodynamical model \cite{Rischke:1995ir,Rischke:1995mt} to simulate
the bulk medium. The heavy-quark propagation in the medium is described
by a relativistic Langevin approach \cite{Rapp:2009my}. 
Similar studies have recently been
performed in a thermal fireball model with a combined
coalescence-fragmentation approach
\cite{vanHees:2007me,vanHees:2007mf,Greco:2007sz,vanHees:2008gj,Rapp:2008fv,
  Rapp:2008qc,Rapp:2009my}, in an ideal hydrodynamics model with a
lattice-QCD EoS \cite{He:2012df,He:2012xz}, in a model from Kolb and
Heinz \cite{Aichelin:2012ww}, in the BAMPS model
\cite{Uphoff:2011ad,Uphoff:2012gb}, the MARTINI model
\cite{Young:2011ug} as well as in further studies and model comparisons
\cite{Moore:2004tg,Vitev:2007jj,Gossiaux:2010yx,Gossiaux:2011ea,Gossiaux:2012th}.

The use of the UrQMD hybrid model provides a major step forward as
compared to simplified expanding fireball models employed so far. It
provides a realistic and well established background, including
event-by-event fluctuations and has been shown to describe well 
many collective properties of relativistic heavy-ion collisions. For the
heavy-quark propagation we apply a Langevin approach.  Within this
framework we investigate the effects of using different drag and
diffusion coefficients and different freeze-out temperatures of heavy
flavors on the heavy-quark observables and compare the results with the
experimental data from the Relativistic Heavy Ion Collider (RHIC) 
and the Large Hadron Collider (LHC).

\section{The UrQMD hybrid model}

To extract information on the interaction of heavy quarks with the
medium one ideally applies a well tested model for the
(collective) dynamics of the bulk matter. In heavy-ion collisions the
medium is by no means homogeneous. Rather it is a locally and
event-by-event fluctuating, fast expanding system. In our calculation we
employ the state of the art UrQMD hybrid model for the description of
the expanding background. This model has been developed in the past years to
combine the advantages of hadronic transport theory and ideal fluid
dynamics \cite{Petersen:2008dd}. To account for the non-equilibrium
dynamics in the very early stage of the collision in the hybrid model,
the UrQMD cascade \cite{Bass:1998ca,Bleicher:1999xi} is used to
calculate the initial states of the heavy ion collisions each to be used
in a subsequent hydrodynamical evolution \cite{Steinheimer:2007iy}. The
transition from the UrQMD initial state and the hydrodynamical evolution
takes place at a time
$t_{\text{start}} = 2R/\sqrt{\gamma_{\text{CM}}^2 -1}$
i.e., after the two Lorentz-contracted nuclei have passed through each
other ($\gamma_{\text{CM}}$ is the center-of-mass-frame Lorentz
factor, and $R$ is the radius of the nucleus). The energy, baryon
number, and momenta of all particles within UrQMD are mapped onto a
spatial grid for the hydrodynamic evolution including event-by-event
fluctuations.  The full (3+1)-dimensional ideal hydrodynamic evolution
is performed using the SHASTA algorithm
\cite{Rischke:1995ir,Rischke:1995mt}. We solve the equations for the
conservation of energy and momentum and for the conservation of the
baryonic charge. With $T^{\mu \nu}$ denoting the relativistic
energy-momentum tensor the corresponding equations read
\begin{equation}
	\partial_{\mu}T^{\mu \nu}= 0,
\end{equation}
and for the baryon four-current $N^{\mu}$
\begin{equation}\label{densprop}
	\partial_{\mu} N^{\mu} = 0.
\end{equation}
To transfer all particles back into the UrQMD model, an approximate
iso-eigentime transition is chosen (see \cite{Li:2008qm} for
details). Here, we apply the Cooper-Frye prescription
\cite{Cooper:1974mv} and transform to particle degrees of freedom via
\begin{equation}\label{cooper}
	E \frac{\dd N}{d^3p} = g_i \int_{\sigma} \dd \sigma_{\mu} \; p^{\mu}
        f(x,p).
\end{equation}
Here $\dd \sigma_{\mu}=(\dd^3x,0,0,0)$) is the hypersurface normal. In
Eq. (\ref{cooper}) $f(x,p)$ are the Bose- and Fermi-distribution functions
and $g_i$ the degeneracy factors for the different particle species.
After the ``particlization'' the evolution proceeds in the hadronic cascade
(UrQMD), where final re-scatterings and decays are calculated until all
interactions cease and the system decouples.

A more detailed description of the hybrid model including parameter
tests and results can be found in \cite{Petersen:2008dd}. 
A comparison to the results employing the non-approximated hypersurface 
can be found in \cite{Huovinen:2012is}.

\section{Heavy-quark diffusion}
\label{sec:HQ-diffusion}

The diffusion of a ``heavy particle'' in a medium consisting of ``light
particles'' can be described with a Fokker-Planck equation
\cite{Svet88,MS97,vanHees:2004gq,Moore:2004tg,HGR05a,vanHees:2007me,Gossiaux:2008jv}.
Here one approximates the collision term of the corresponding Boltzmann
equation, which in turn can be mapped into an equivalent stochastic
Langevin equation.

\subsection{Relativistic Langevin approach}
\label{subsec:Langevin}

In the relativistic realm such a Langevin process reads
\begin{equation}
\begin{split}
\label{lang.1}
\dd x_j &= \frac{p_j}{E} \dd t, \\
\dd p_j &= -\Gamma p_j \dd t + \sqrt{\dd t} C_{jk} \rho_k.
\end{split}
\end{equation}
Here $\dd t$ is the time step in the Langevin calculation, $\dd x_j$ and
$\dd p_j$ are the coordinate and momentum changes in each time-step,
$E=\sqrt{m^2+\bvec{p}^2}$, and $\Gamma$ is the drag or friction
coefficient. The covariance matrix, $C_{jk}$, of the fluctuating force
is related to the diffusion coefficients, as we shall see below. Both
$\Gamma$ and $C_{jk}$ dependent on $(t,\bvec{x},\bvec{p})$ and are
defined in the (local) rest-frame of the fluid. The $\rho_k$ are
Gaussian-normal distributed random variables. Their distribution
function reads
\begin{equation}
\label{lang.2}
P(\bvec{\rho}) = \left (\frac{1}{2 \pi} \right)^{3/2} \exp
\left(-\frac{\bvec{\rho}^2}{2} \right ).
\end{equation}
with $\bvec{\rho}=(\rho_1,\rho_2,\rho_3)$.
The fluctuating force $F_j^{(\text{fl})}$ thus obeys
\begin{equation}
\label{lang.3}
\erw{F_j^{(\text{fl})}(t)}=0, \quad \erw{F_j^{(\text{fl})}(t)
  F_k^{(\text{fl})}(t')} = C_{jl} C_{kl} \delta(t-t').
\end{equation}
It is important to note that with these specifications the random
process is not yet uniquely determined since one has to specify, at
which momentum argument the covariance matrix $C_{jk}$ has to be taken
to define the stochastic time integral in (\ref{lang.1}). Thus, we set
\begin{equation}
\label{lang.4}
C_{jk} =C_{jk}(t,\bvec{x},\bvec{p}+\xi \dd \bvec{p}).
\end{equation}
For $\xi=0$, $\xi=1/2$, and $\xi=1$ the corresponding Langevin processes
are called the pre-point Ito, the mid-point Stratonovic-Fisk, and the
post-point Ito (or H\"anggi-Klimontovich) realization, respectively
\cite{Dunkel-Haenggi:2008}.

According to (\ref{lang.1}) and (\ref{lang.3}), for a given value of
$\xi$ in (\ref{lang.4}) the average of an arbitrary observable
$g(\bvec{x},\bvec{p})$ obeys the time-evolution
\begin{equation}
\begin{split}
\label{lang.5}
\erw{g(\bvec{x}+\dd \bvec{x},\bvec{p}+\dd \bvec{p})-g(\bvec{x},\bvec{p})}
  = \Bigg \langle & \frac{\partial g}{\partial x_j} \frac{p_j}{E} +
  \frac{\partial g}{\partial p_j} \left (-\Gamma p_j+\xi \frac{\partial
      C_{jk}}{\partial p_l} C_{lk} \right ) \\
& + \frac{1}{2} \frac{\partial^2 g}{\partial p_j \partial p_k} C_{jl}
C_{kl} \Bigg \rangle \dd t + \mathcal{O}(\dd t^{3/2}).
\end{split}
\end{equation}
Here all momentum arguments of the drag and diffusion coefficients have
to be taken at $\bvec{p}$. From (\ref{lang.5}) it follows immediately
that the time evolution of the phase-space distribution function
$f_Q(t,\bvec{x},\bvec{p})$ of heavy quarks is given by the Fokker-Planck
equation,
\begin{equation}
\label{lang.6}
\frac{\partial f_Q}{\partial t} + \frac{p_j}{E} \frac{\partial f_Q}{\partial x_j} =
\frac{\partial}{\partial p_j} \left [ \left (\Gamma p_j-\xi C_{lk}
    \frac{\partial C_{jk}}{\partial p_l} \right ) f_Q \right ] +
\frac{1}{2} \frac{\partial^2}{\partial p_j \partial p_k} (C_{jl} C_{kl} f_Q).
\end{equation}
Thus, the usual drag and diffusion coefficients for an isotropic medium
are related to the pertinent parameters in the Langevin process by
\begin{alignat}{2}
\label{lang.7}
  A p_j &=\Gamma p_j - \xi C_{lk} \frac{\partial C_{jk}}{\partial p_l},
  \\
\label{lang.8}
  C_{jk} &=\sqrt{2 B_0} P_{jk}^{\perp} + \sqrt{2 B_1} P_{jk}^{\perp}, \\
\label{lang.9}
  \text{with} \quad P_{jk}^{\parallel} &=\frac{p_j p_k}{\bvec{p}^2}, \quad
  P_{jk}^{\perp}=\delta_{jk}-\frac{p_j p_k}{\bvec{p}^2}.
\end{alignat}
In\footnote{In numerical studies it has turned out that drag and
  diffusion coefficients as obtained from microscopic models usually do
  not lead to the expected long-time stationary limit of the phase-space
  distribution for the heavy particles when diffusing in an equilibrated
  background medium.} case of a homogeneous static background (``heat
bath''), the stationary limit should become a Boltzmann-J\"uttner
distribution with the temperature of the ``heat bath''. Thus, one
typically adjusts the drag coefficient by choosing the longitudinal
diffusion coefficient, $B_1$, in (\ref{lang.8}) such as to satisfy this
asymptotic equilibration condition, leading to dissipation-fluctuation
relations between this diffusion coefficient and the drag coefficient
\cite{Moore:2004tg,Rapp:2009my}.

It turns out that for $B_0=B_1=D(E)$ and a homogeneous background medium
the Boltzmann-J\"uttner distribution,
\begin{equation}
\label{lang.10}
f_Q^{(\text{eq})}(\bvec{p})=\exp \left(-\frac{E}{T} \right ), \quad
\text{with} \quad E=\sqrt{\bvec{p}^2+m^2},
\end{equation}
becomes a solution of the corresponding stationary Fokker-Planck
equation, if the dissipation-fluctuation relation
\begin{equation}
\label{lang.11}
\Gamma(E) E T-D(E)+ T(1-\xi) D'(E)=0,
\end{equation}
is fulfilled. A straightforward way to achieve the correct asymptotic
equilibrium distribution within a relativistic Langevin simulation is to
set $\xi=1$ (i.e., using the post-point Ito realization). This reduces
(\ref{lang.11}) to
\begin{equation}
\label{lang.12}
D(E)=\Gamma(E) E T.
\end{equation}
For applications to heavy-ion collisions we use $\Gamma$ and $B_0$ from
underlying microscopic models for heavy-quark scattering with light
quarks and gluons as detailed below and adjust the longitudinal
diffusion coefficient to
\begin{equation}
\label{lang.13}
B_1=\Gamma E T.
\end{equation}
So far we have defined our Langevin process with respect to the (local)
rest frame of the background medium. For a medium with collective flow,
one has to evaluate the time-step in the local rest-frame and boost-back
to the computational frame.  For a closer look on the post-point
description see section \ref{AppA}.

\begin{figure}[h]
\begin{minipage}[b]{0.45\textwidth}
\includegraphics[width=1\textwidth]{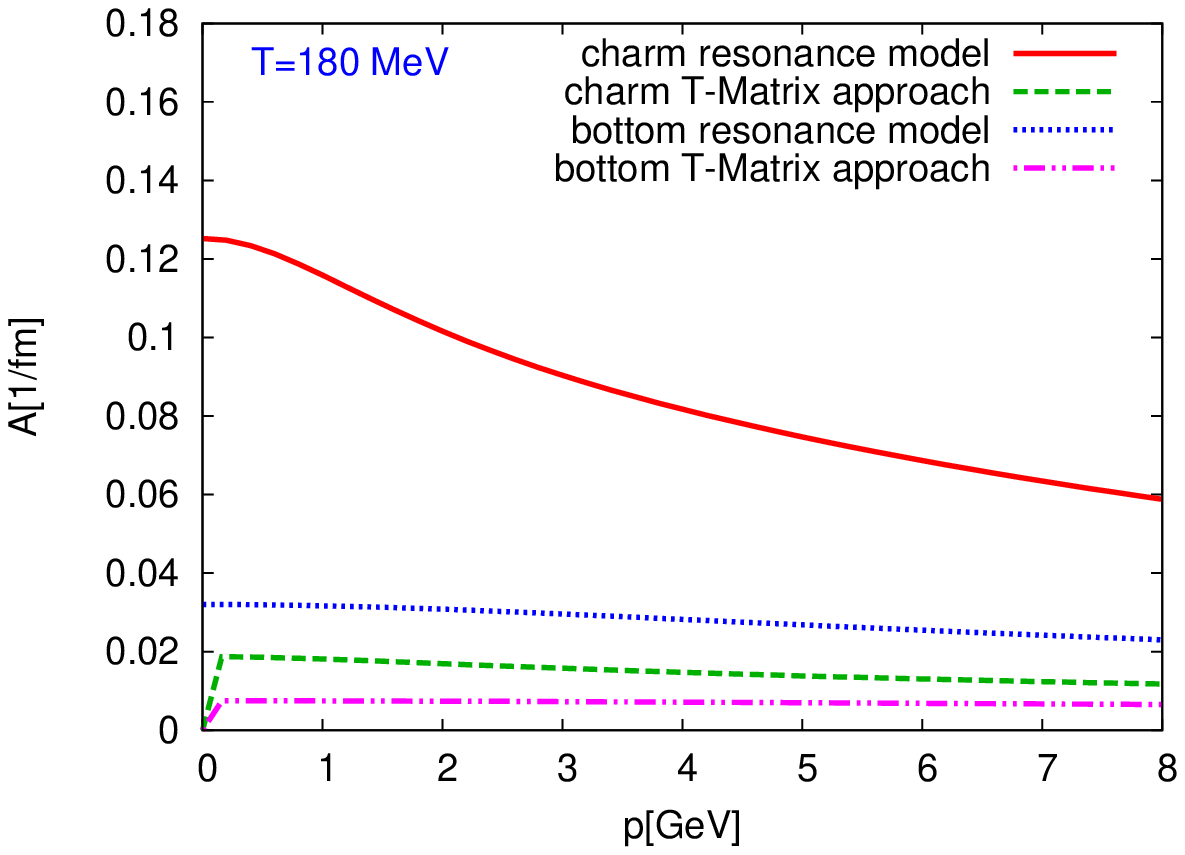}
\end{minipage}
\hspace{5mm}
\begin{minipage}[b]{0.45\textwidth}
\includegraphics[width=1\textwidth]{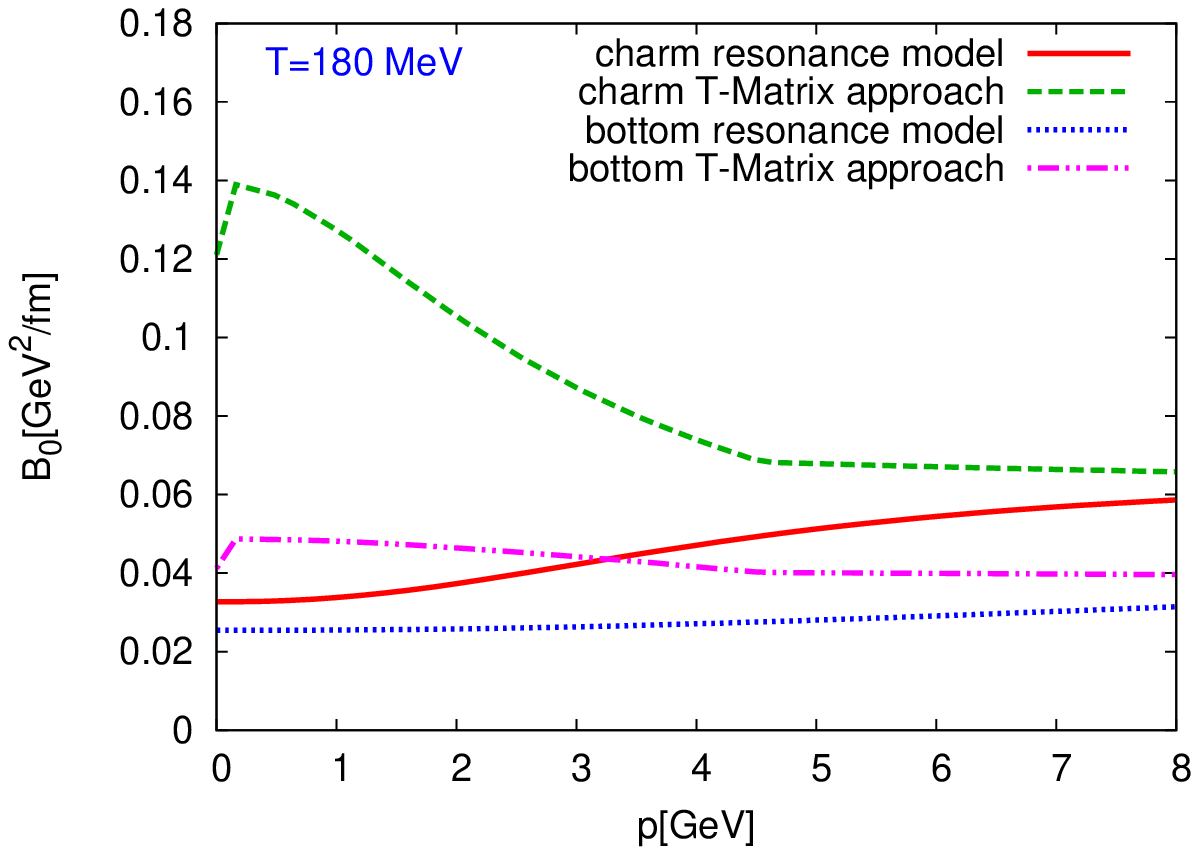}
\end{minipage}
\caption{(Color online) Drag (left) and diffusion (right) coefficients in the resonance
  model and the T-Matrix approach for charm and bottom quarks.  The plot
  shows the dependence of the coefficients on the three-momentum
  $|\vec{p}|$ at a fixed temperature of $T=180\,\text{MeV}$.  }
\label{Coeffp}
\end{figure}

\begin{figure}[h]
\begin{minipage}[b]{0.45\textwidth}
\includegraphics[width=1\textwidth]{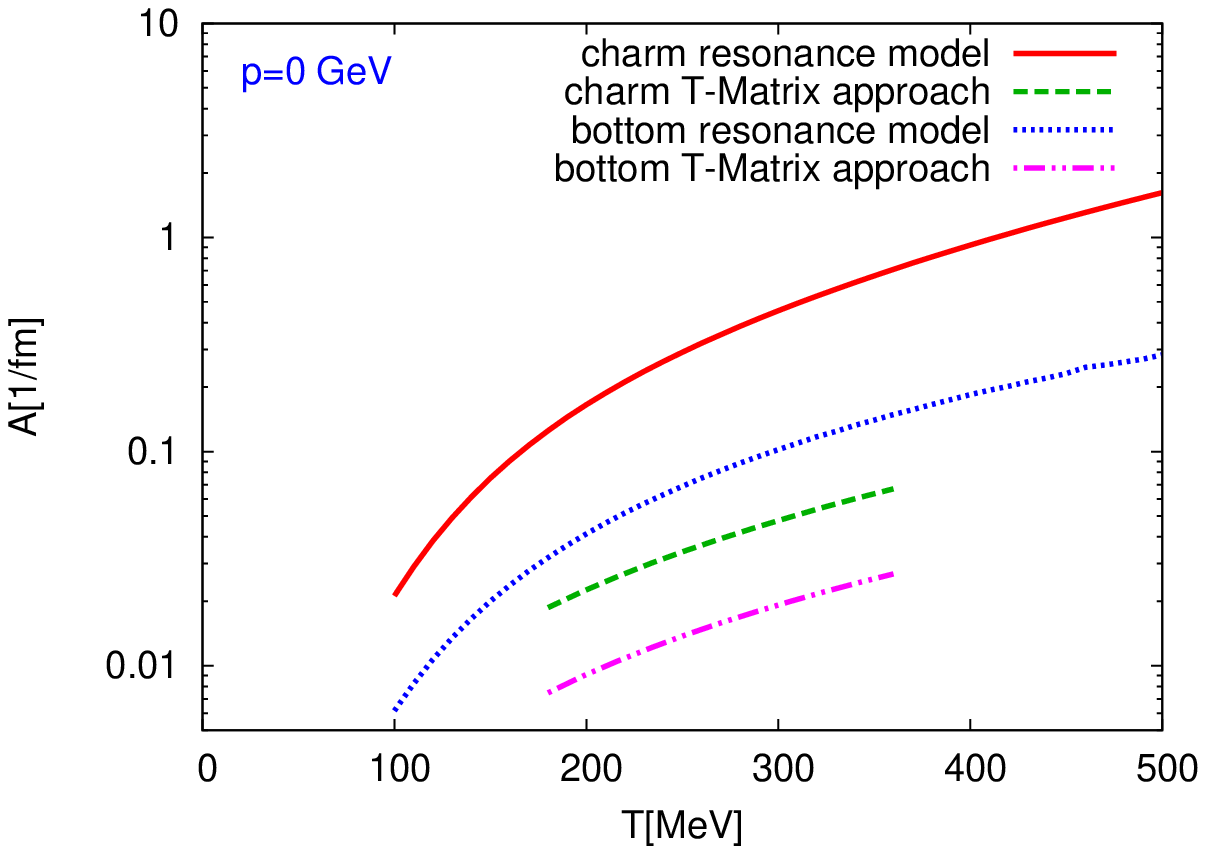}
\end{minipage}
\hspace{5mm}
\begin{minipage}[b]{0.45\textwidth}
\includegraphics[width=1\textwidth]{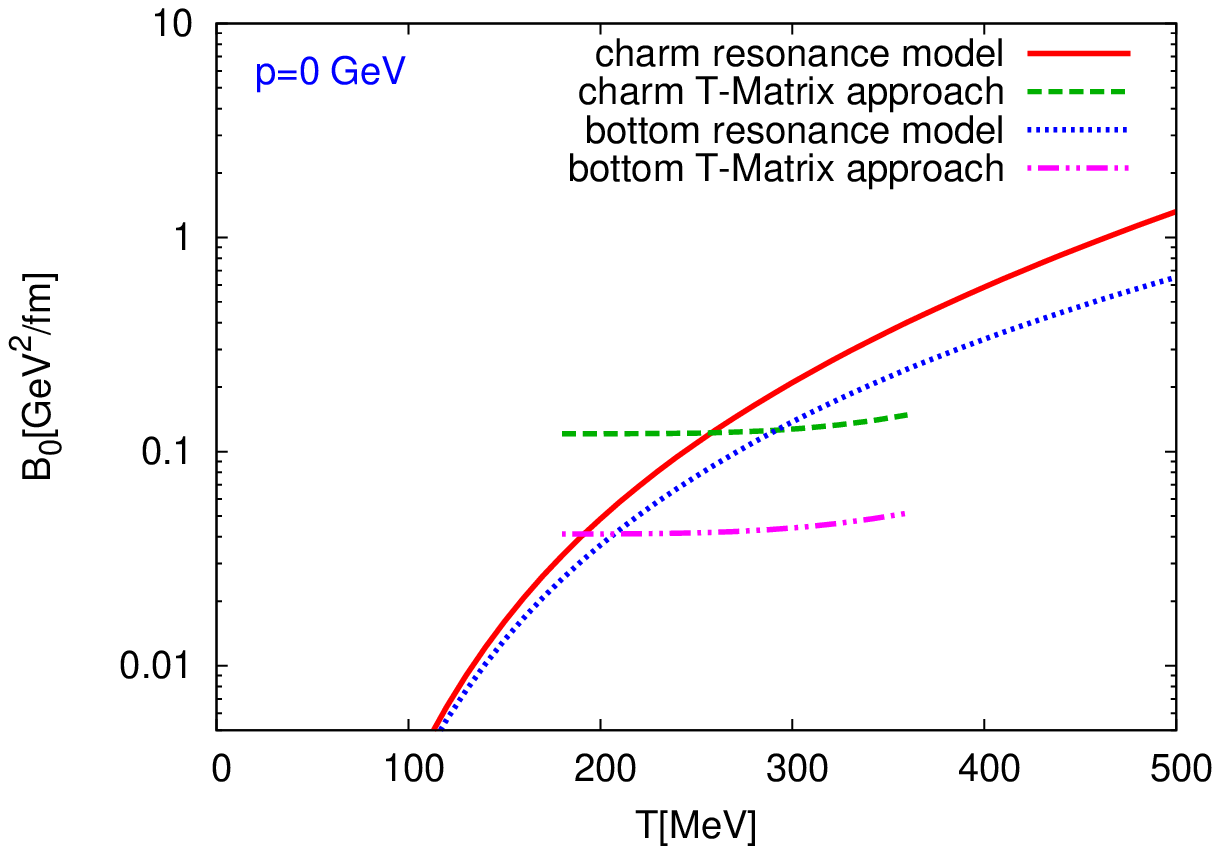}
\end{minipage}
\caption{(Color online) Drag (left) and diffusion (right) coefficients in the resonance
  model and the T-Matrix approach for charm and bottom quarks.  The plot
  shows the dependence of the coefficients on the temperature at a fixed
  three-momentum $|\vec{p}|=0$. 
  The T-Matrix coefficients are calculated between $180\,\text{MeV}$ and $360\,\text{MeV}$ only. }
\label{CoeffT}
\end{figure}

For the heavy-quark propagation in the Langevin model we also need
transport coefficients. In this work these drag and diffusion
coefficients are obtained from two non-perturbative models for elastic
heavy-quark scattering, a resonance model, where the existence of
D-mesons and B-mesons in the QGP phase is assumed, as well as a
$T$-Matrix approach in which quark-antiquark potentials are used for the
calculation of the coefficients in the QGP.  They are described in
detail in Sec. \ref{AppB} and are shown in Fig. \ref{Coeffp} as function of
the three-momentum $|\vec{p}|$ at $T=180\,\text{MeV}$ and in Fig.
\ref{CoeffT} as function of the temperature at a fixed three-momentum of
$|\vec{p}|=0$.

\subsection{Implementation of the Langevin simulation into the
  UrQMD-hybrid model}

For the present study, charm production and propagation is evaluated
perturbatively on the time-dependent background generated by
UrQMD/Hybrid. To model a fluctuating and space-time dependent
Glauber-initial state geometry, we perform a first UrQMD run with
elastic $0^\circ$ scatterings between the colliding nuclei and save the
nucleon-nucleon collision space-time coordinates. These coordinates are
used in a second, full UrQMD run as (possible) production coordinates
for the heavy quarks.

\begin{figure}[h]
\begin{minipage}[b]{0.45\textwidth}
\includegraphics[width=1\textwidth]{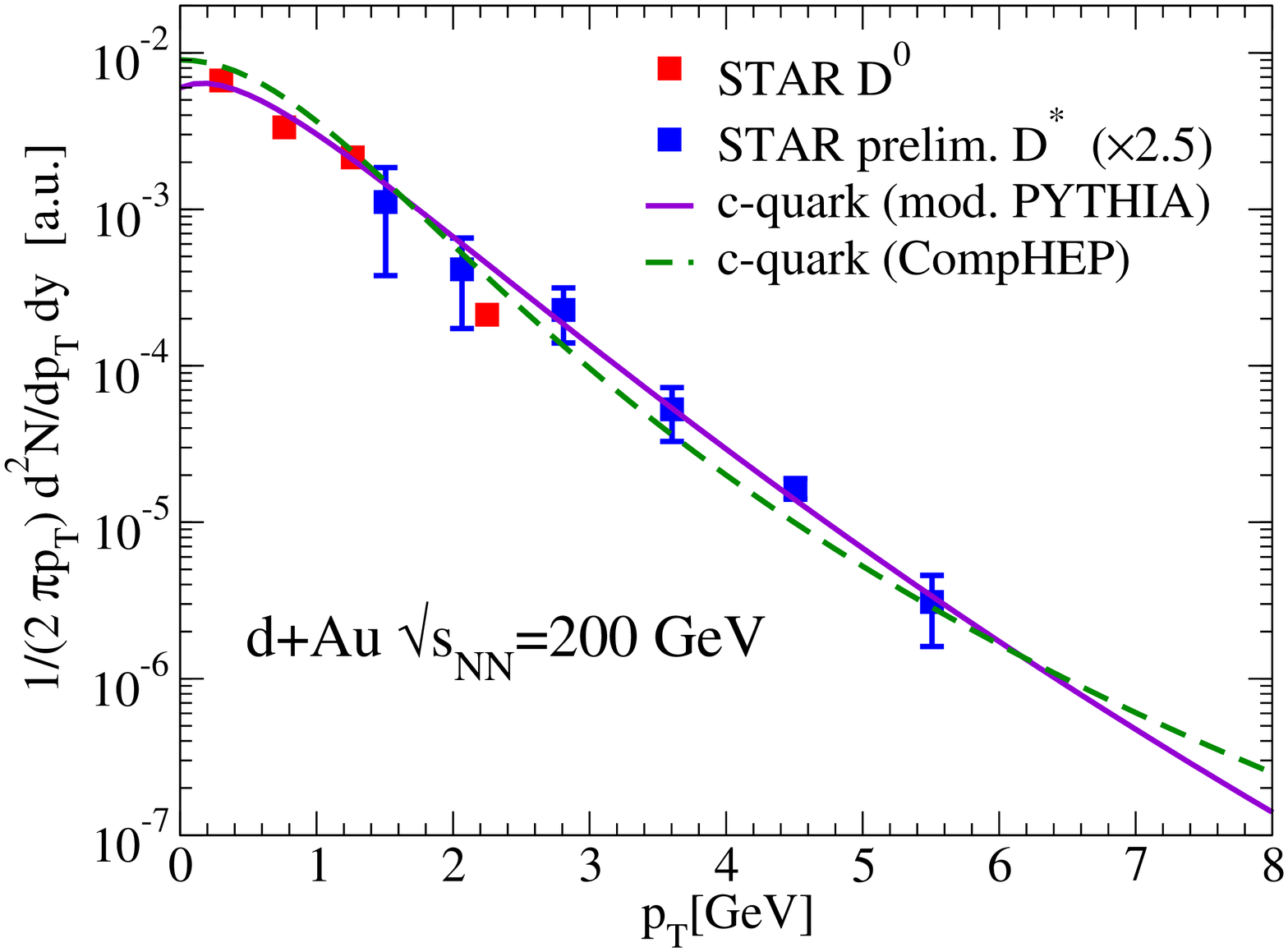}
\end{minipage}
\hspace{7mm}
\begin{minipage}[b]{0.45\textwidth}
\includegraphics[width=1\textwidth]{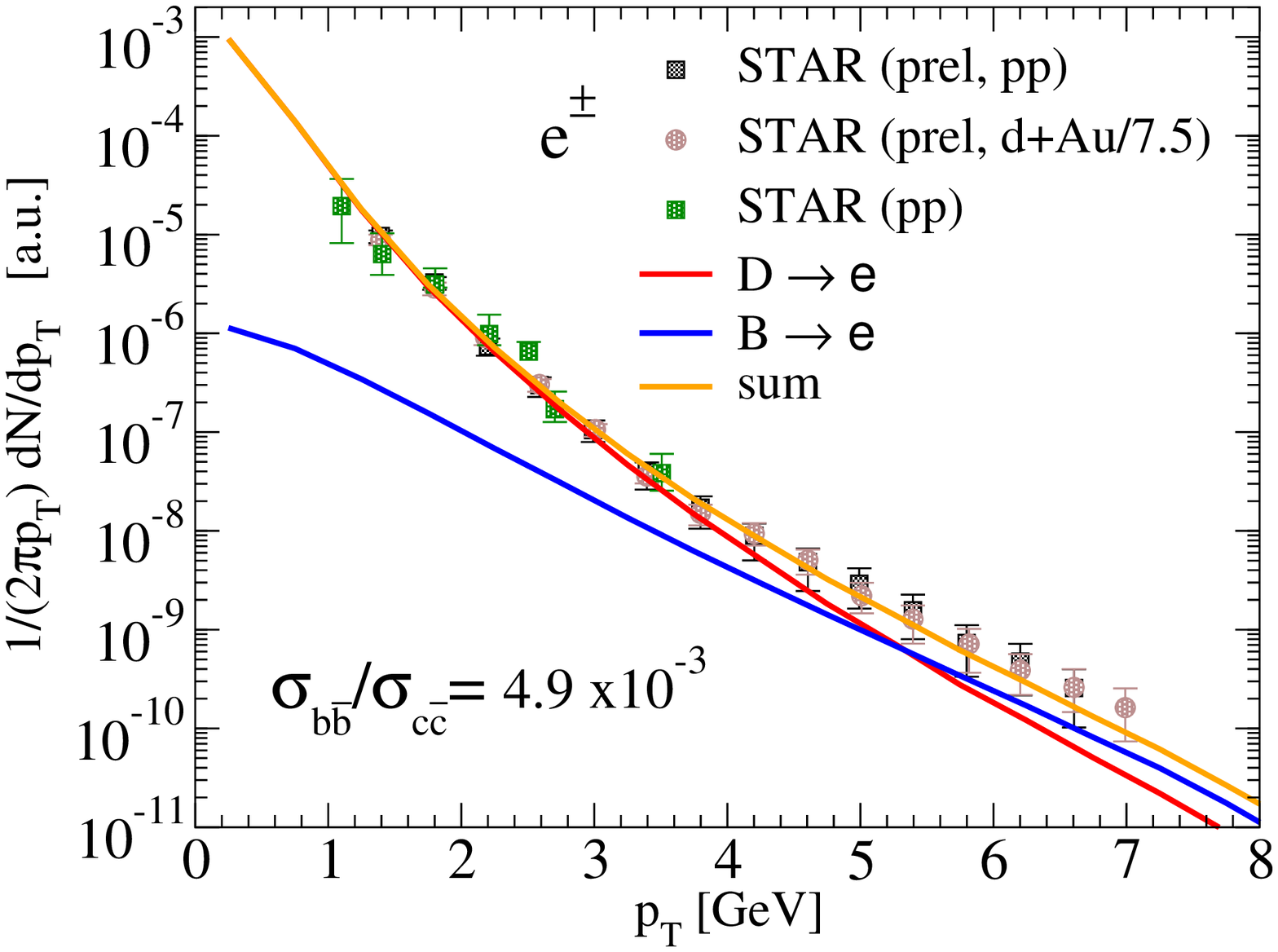}
\end{minipage}
\caption{(Color online) Fits of $D$- and $D^*$ -meson $p_T$ spectra in
  $200\,\text{AGeV}$ d-Au collisions at RHIC with a modified PYTHIA
  simulation (left panel) and the corresponding non-photonic
  single-electron $p_T$ spectra in p-p and d-Au collisions (taken from 
  \cite{Rapp:2005at}). The missing yield of high-$p_T$ electrons is
  fitted with the analogous B-meson decay spectra, thus fixing the
  bottom-charm ratio at $\sigma_{b\bar{b}} /\sigma_{c\bar{c}} \simeq
  4.9\cdot 10^{-3}$.}
\label{dist}
\end{figure}

As momentum distribution for the initially produced charm quarks at
$\sqrt {s_{NN}}=200\; \GeV$ we use
\begin{equation}
\frac{1}{2\pi p_Tdp_T}=\frac{\left(A_1+p_T^2\right)^2}{\left(1+A_2\cdot p_T^2\right)^{A_3}},
\end{equation}
with $A_1=0.5$, $A_2=0.1471$ and $A_3=21$ and for bottom
quarks
\begin{equation}
\frac{1}{2\pi p_Tdp_T}=\frac{1}{\left( A_1+p_T^2 \right)^{A_2}},
\end{equation}
with $A_1=57.74$ and $A_2=5.04$.  These distributions are
taken from \cite{vanHees:2005wb,vanHees:2007me} and are shown in
Fig. \ref{dist}. They are obtained by using tuned c-quark spectra from
PYTHIA.  Their pertinent semileptonic single-electron decay spectra
account for pp and dAu measurements by STAR up to $4\,\text{GeV}$.  The
missing part at higher $p_T$ is then supplemented by B-meson
contributions.

Starting with these charm- and bottom-quark distributions as initial
conditions we perform, as soon as the hydrodynamics start condition is
fulfilled, an Ito post-point time-step of our Langevin simulation as
described in Sec.~\ref{subsec:Langevin}, at each time-step of the
hydrodynamical evolution.

We use the cell velocities, cell temperatures, the length of the
time-step and the $\gamma$-factor of the cells to calculate the momentum
transfer, propagating all heavy quarks independently. For the Langevin
transport we use the drag and diffusion coefficients obtained from
the resonance model or $T$-Matrix approach as described in
Sec.~\ref{AppB}.

To analyze the sensitivity of $R_{AA}$ and especially $v_2$ on the
decoupling time of the heavy flavors from the medium we vary the
decoupling temperatures between $130 \; \MeV$ and $180\; \MeV$ (for the
resonance model) and extrapolate the corresponding transport
coefficients smoothly into the hadronic phase.  This assumption of a
smooth transition of the transport coefficients in the transition from
the partonic description above and the hadronic one below $T_c$ has been
verified, using an effective model for open-heavy-flavor interactions in
a hadronic medium in \cite{He:2011yi,He:2012df}.
 
Our approach provides us with the heavy-quark momentum distribution. We
include a hadronization mechanism for open-heavy-flavor mesons (D and B
mesons). Since non-photonic single electrons are usually measured in experiments, we 
perform a semileptonic decay into electrons as final step to compare to
data. In addition we also provide D- and B-meson results for direct
comparisons to the upcoming direct D/B measurements by the STAR Heavy
Flavor Tracker (HFT). These results are shown in Sec.~\ref{AppC}.

\section{Results at RHIC energies}

\subsection{Elliptic flow $v_2$ and nuclear modification factor $R_{AA}$ with fragmentation}

\begin{figure}[h]
\begin{minipage}[b]{0.45\textwidth}
\includegraphics[width=1\textwidth]{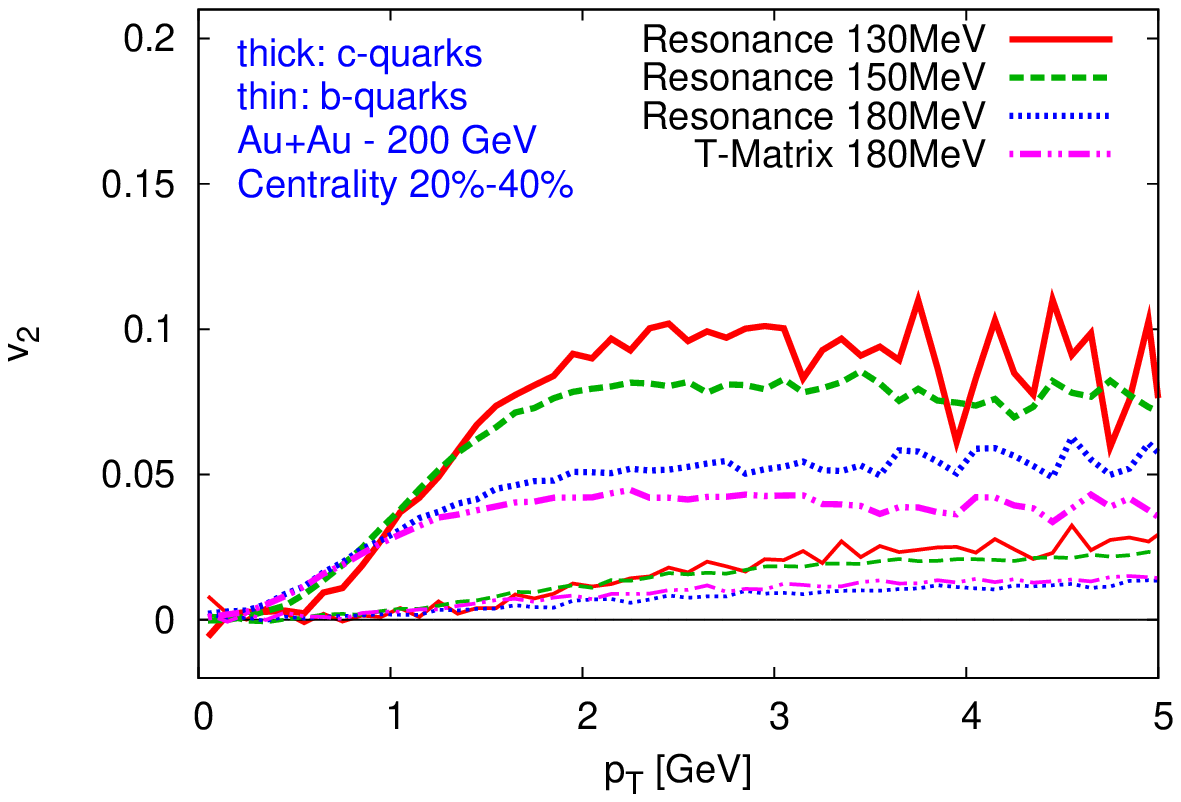}
\end{minipage}
\begin{minipage}[b]{0.45\textwidth}
\includegraphics[width=1\textwidth]{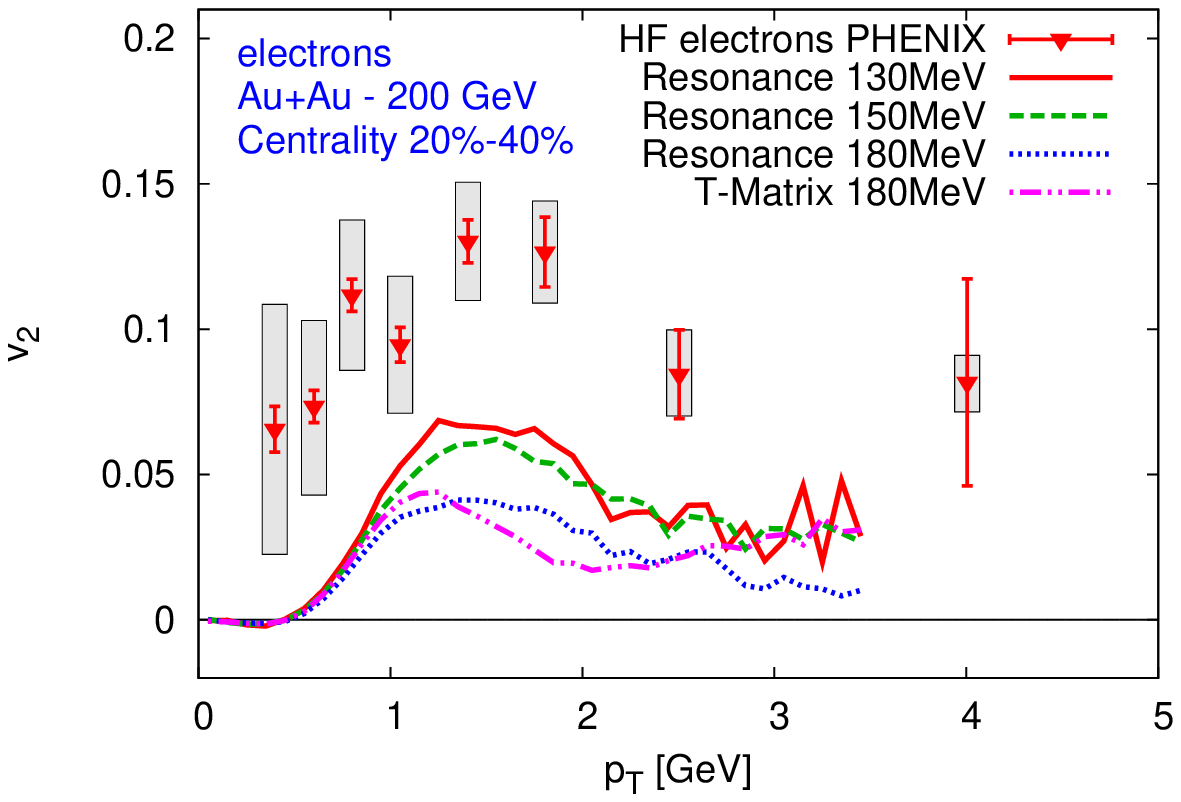}
\end{minipage}
\caption{(Color online) Left: Elliptic flow, $v_2$, of heavy quarks in
  Au+Au collisions at $\sqrt {s_{NN}}=200\; \GeV$. We use a rapidity cut
  of $|y|<0.35$. The solid lines depict the charm quarks while the
  dashed lines depict the bottom quarks. \\
  Right: Elliptic flow, $v_2$, of electrons from heavy-meson decays
  using Peterson fragmentation to D/B mesons and subsequent decay into
  electrons in Au+Au collisions at $\sqrt {s_{NN}}=200\; \GeV$.  We use
  a rapidity cut of $|y|<0.35$. Data are from \cite{Adare:2010de}.}
\label{flowRHIC4}
\end{figure}

Fig.~\ref{flowRHIC4} presents the elliptic flow, $v_2$, of charm and
bottom quarks from Au+Au collisions at $\sqrt{s_{NN}} =200\;\text{GeV}$
in the centrality range $\sigma /\sigma_{tot}=$20\%-40\% applying a
rapidity cut of $|y|<0.35$.

For our calculation using the drag and diffusion coefficients of the
$T$-Matrix model we use a decoupling temperature of $180\;\MeV$, while
with the resonance model we show results for decoupling temperatures of
$130\;\MeV$, $150\;\MeV$ and $180\;\MeV$.

As one can clearly see, the elliptic flow, $v_2$, of bottom quarks
(dashed lines) is much smaller compared to that of the charm quarks
(solid lines) due to their larger mass. Furthermore the use of the
coefficients from the $T$-Matrix model compared with those from the
resonance model shows that both calculations are in reasonable
agreement.  The elliptic flow of the charm quarks is nevertheless
somewhat lower for the $T$-Matrix model than for the resonance
model. When decreasing the decoupling temperature the flow clearly
increases. Thus, we conclude that the late phase of the heavy-ion
collision may have considerable influence on the heavy-flavor elliptic
flow although the drag and diffusion coefficients become small in the
late stages of the fireball evolution.

Moreover the, $v_2$, is shifted towards higher $p_T$ for lower
decoupling temperatures. This effect is due to the increased radial
velocity of the medium, which is in case of an developed elliptic flow
larger in $x$ than in $y$ direction. Consequently there is a depletion
of particles with high $v_x$ in the low $p_T$ region and smaller
elliptic flow. This effect is more important for heavier particles and a
larger radial flow \cite{Huovinen:2001cy,Krieg:2007bc}.

To compare our calculations with data on non-photonic electrons from
RHIC we perform a Peterson fragmentation of the charm and bottom quarks
to D-mesons and B-mesons using the fragmentation function from
\cite{Peterson:1982ak},
$$D^H_Q(z)=\frac{N}{z[1-(1/z)-\epsilon_Q/(1-z)]^2},$$
where $N$ is a normalization constant, $z$ the relative-momentum
fraction obtained in the fragmentation of the heavy quark and
$\epsilon_Q=0.05 (0.005)$ for charm (bottom) quarks. After hadronization
we use PYTHIA routines for the semileptonic decay to electrons
\cite{Sjostrand:2006za,Sjostrand:2007gs}.

Fig. \ref{flowRHIC4} shows our results for the $v_2$ for
single-electrons in comparison to data from the PHENIX collaboration.

Again we clearly observe the importance of the late phase of the
collision. The depletion effect at low $p_T$ described before is clearly
visible. The decrease of the elliptic flow at high $p_T$ is due to the
increasing fraction of electrons from bottom decays, which have a lower
$v_2$ as seen in Fig.~\ref{flowRHIC4}.  The calculated flow in the setup
with the Peterson fragmentation is too small compared to the PHENIX
data.

\begin{figure}[h]
\center
\begin{minipage}[b]{0.45\textwidth}
\includegraphics[width=1\textwidth]{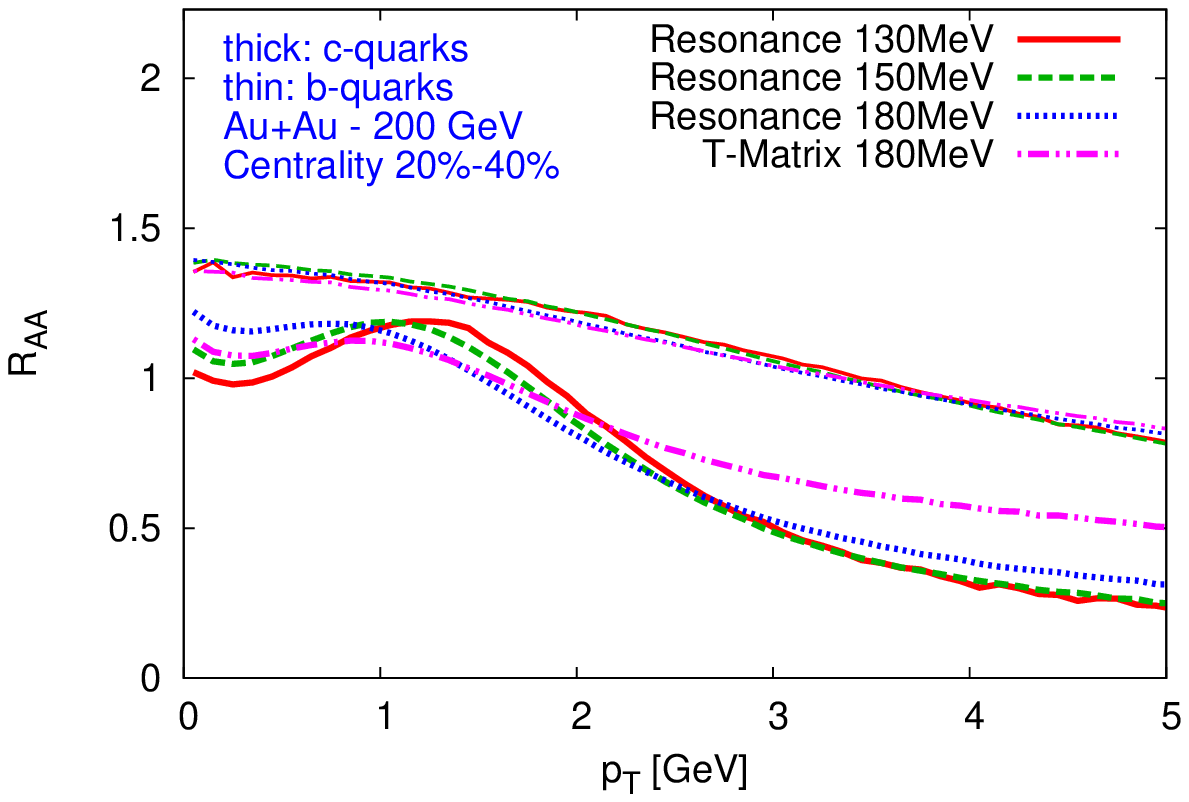}
\end{minipage}
\begin{minipage}[b]{0.45\textwidth}
\includegraphics[width=1\textwidth]{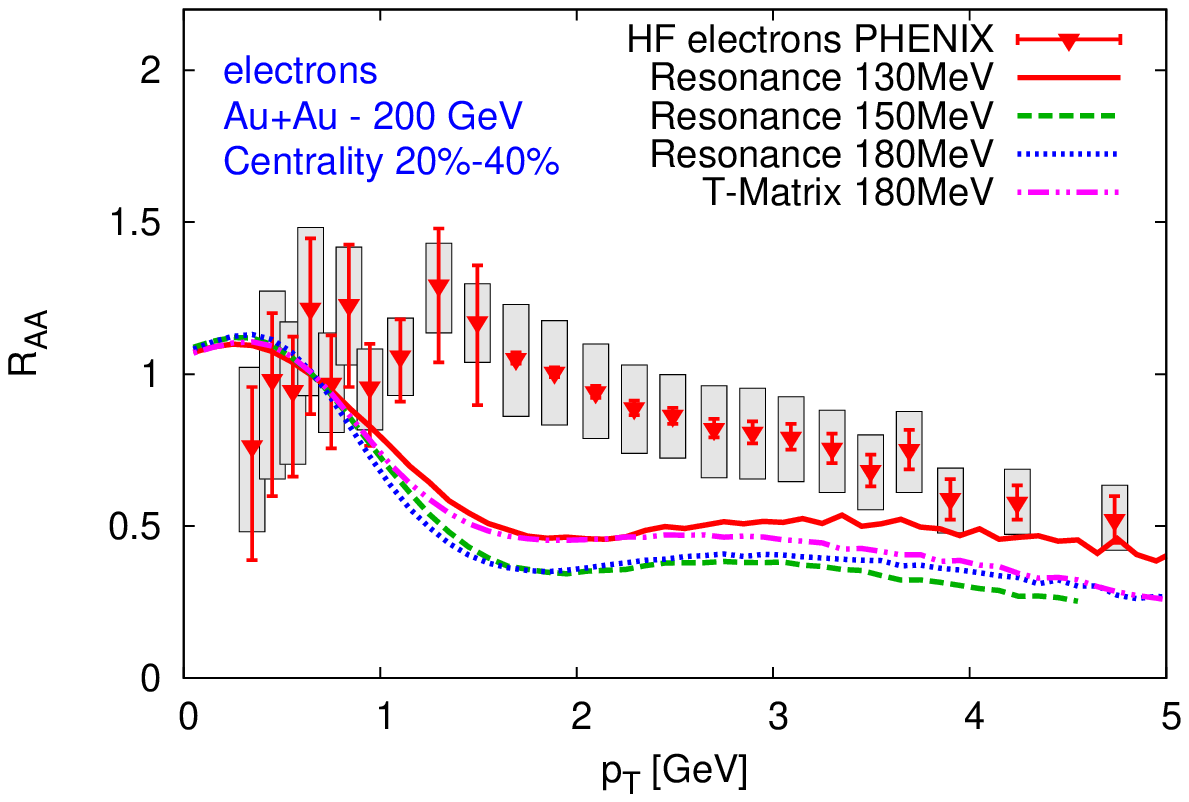}
\end{minipage}
\caption{(Color online) Left: $R_{AA}$ of heavy quarks in Au+Au
  collisions at $\sqrt {s_{NN}}=200\;\GeV$.  We use a rapidity cut of
  $|y|<0.35$. The solid lines depict the charm quarks while the dashed
  lines depict the
  bottom quarks.\\
  Right: $R_{AA}$ of electrons from heavy quark decays in Au+Au
  collisions at $\sqrt {s_{NN}}=200\;\GeV$ compared to RHIC data
  \cite{Adare:2010de}. We use a rapidity cut of $|y|<0.35$. The
  high-$p_T$ suppression turns out to be too strong compared with the
  data.}
\label{RAARHIC4}
\end{figure}

The corresponding nuclear modification factor, $R_{AA}$, for heavy
quarks is shown in Fig. \ref{RAARHIC4}.  Again we present results for
Au+Au collisions at $\sqrt{s_{NN}} =200\;\GeV$ in the centrality range
20\%-40\%. The quenching for charm quarks is, as expected, much stronger
than for bottom quarks\footnote{However, recent preliminary PHENIX data
  presented at QM 2012 seem to suggest the contrary.\\(Talk by T.~Sakaguchi at QM 2012, data not published yet)}.  While for bottom
quarks the suppression at high $p_T$ is moderate, $R_{AA}$ may drop to
20-30\% for charm quarks. The influence of the medium is, as already
seen in our flow calculations, larger for a lower decoupling temperature
underlining the importance of the late phase of the
collision. Fig.~\ref{RAARHIC4} shows the comparison of our
non-photonic-electron $R_{AA}$ to the data taken by the PHENIX
collaboration.

The nuclear modification factor drops quite rapidly and stabilizes at
about $p_T\gtrsim 2\;\GeV$. Around $p_T\approx 2\;\GeV$ it is
significantly below the PHENIX data. For higher $p_T$ the calculated
$R_{AA}$ approaches the measured data, especially for low decoupling
temperatures. This effect is due to the increasing flow of the
heavy-flavor particles with decreasing decoupling temperature, which
pushes low-$p_T$ heavy-flavor particles towards higher $p_T$ bins.

\subsection{Elliptic flow $v_2$ and nuclear modification factor $R_{AA}$ using a $k$ factor}
 
In the previous section we learned that the elliptic flow of the heavy
quarks in the calculation with fragmentation is too small compared to experimental data.
One possibility to improve on this problem may be to multiply the drag and
diffusion coefficients with a ``$k$ factor''.  Therefore we have
performed the same calculations as in the last section but using a $k$
factor of 3.

\begin{figure}[h]
\begin{minipage}[b]{0.45\textwidth}
\includegraphics[width=1\textwidth]{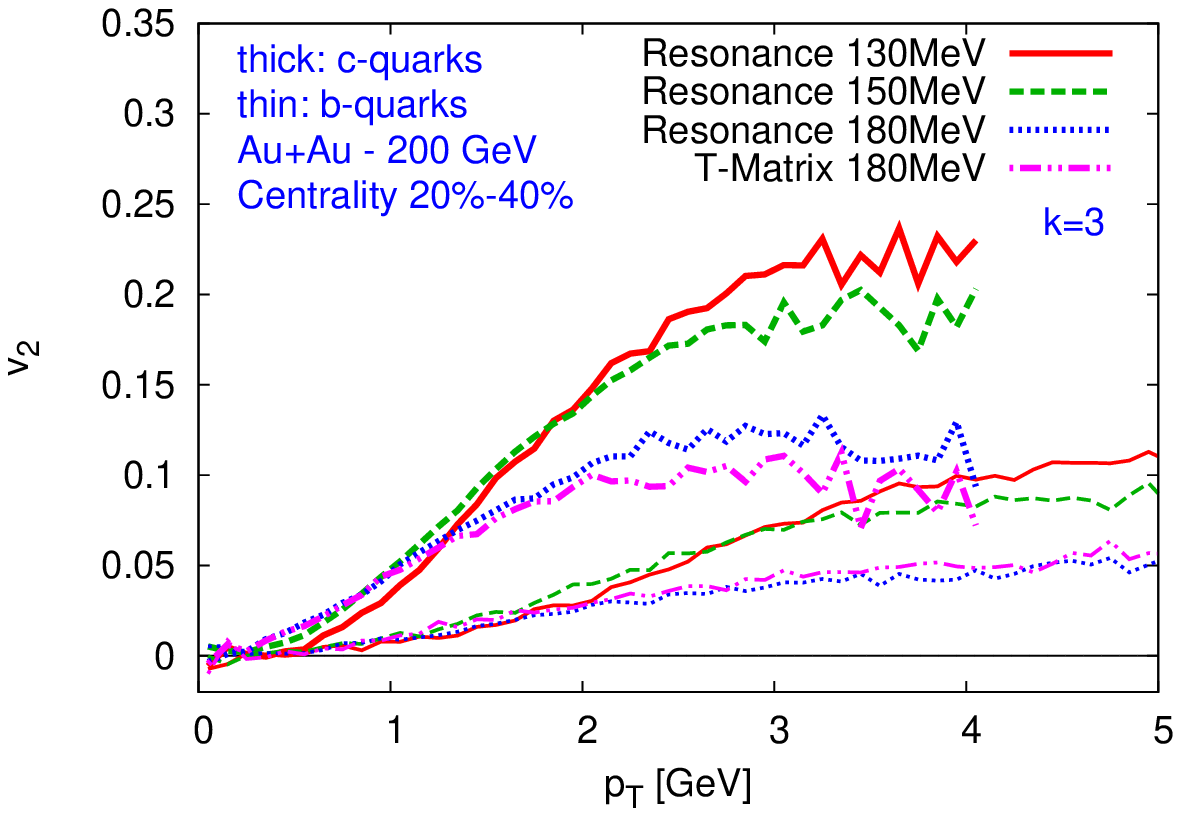}
\end{minipage}
\begin{minipage}[b]{0.45\textwidth}
\includegraphics[width=1\textwidth]{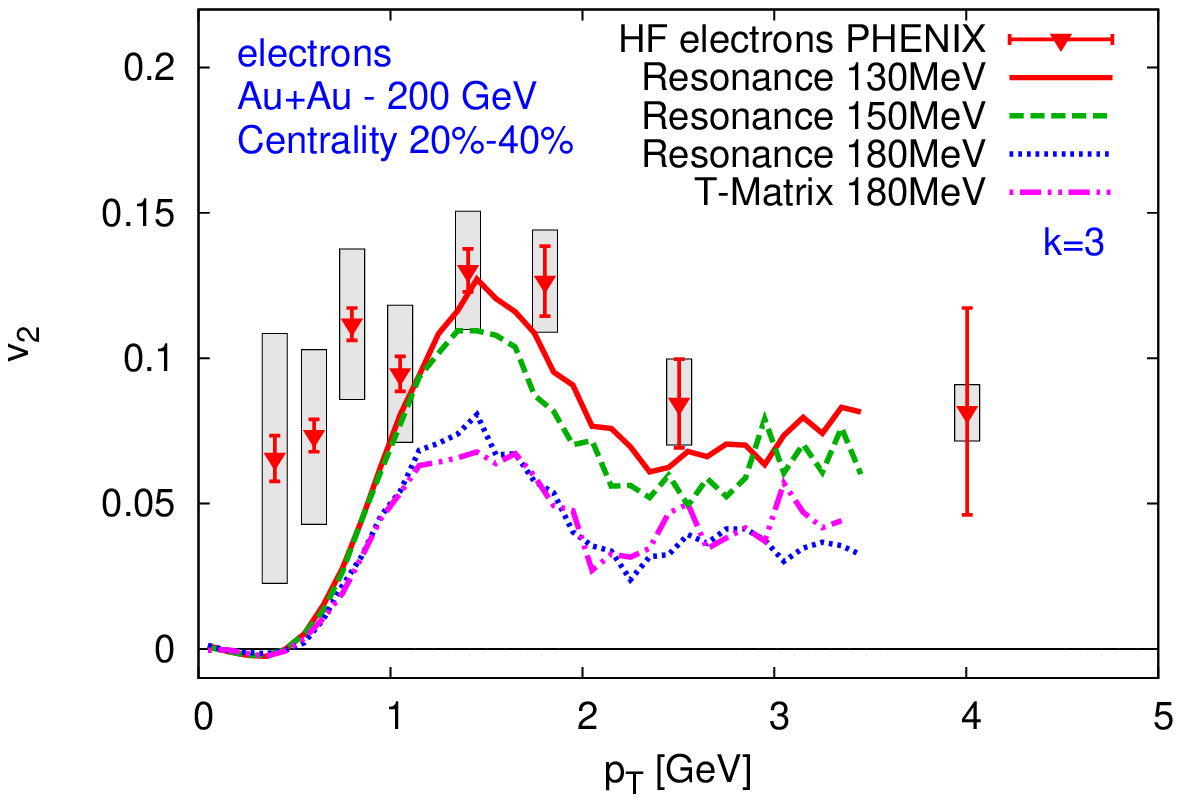}
\end{minipage}
\caption{(Color online) Left: Elliptic flow, $v_2$, of heavy quarks in
  Au+Au collisions at $\sqrt {s_{NN}}=200\;\GeV$ employing a $k$ factor
  of 3. We use a rapidity cut of $|y|<0.35$. The solid lines depict the
  charm quarks while the dashed lines depict the bottom quarks.\\
  Right: Elliptic flow, $v_2$, of electrons from heavy quark decays in
  Au+Au collisions at $\sqrt {s_{NN}}=200\;\GeV$ employing a $k$ factor
  of 3. We use a rapidity cut of $|y|<0.35$. The flow in our calculation
  using a $k$ factor is comparable to data \cite{Adare:2010de}.}
\label{k3flowRHIC4}
\end{figure}

As we see in Fig.~\ref{k3flowRHIC4} the elliptic flow increases
considerably due to the stronger coupling of the heavy quarks to the hot
medium. The results after performing the Peterson fragmentation and the
subsequent decays to electrons are shown in Fig. \ref{k3flowRHIC4}.
The elliptic flow is now comparable to the data, especially when using a
low decoupling temperature of $130\;\MeV$.  Only at low $p_T$ we
underestimate the flow due to the depletion effect described above.

\begin{figure}[h]
\begin{minipage}[b]{0.45\textwidth}
\includegraphics[width=1\textwidth]{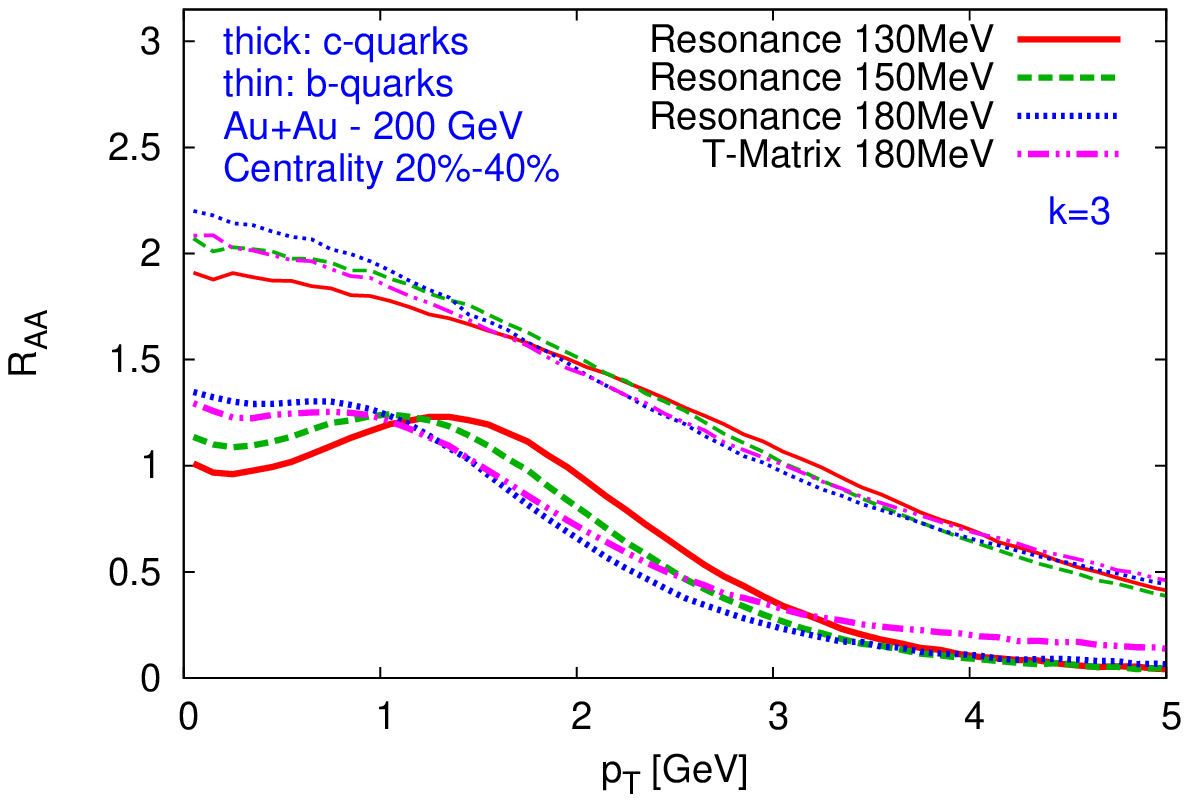}
\end{minipage}
\begin{minipage}[b]{0.45\textwidth}
\includegraphics[width=1\textwidth]{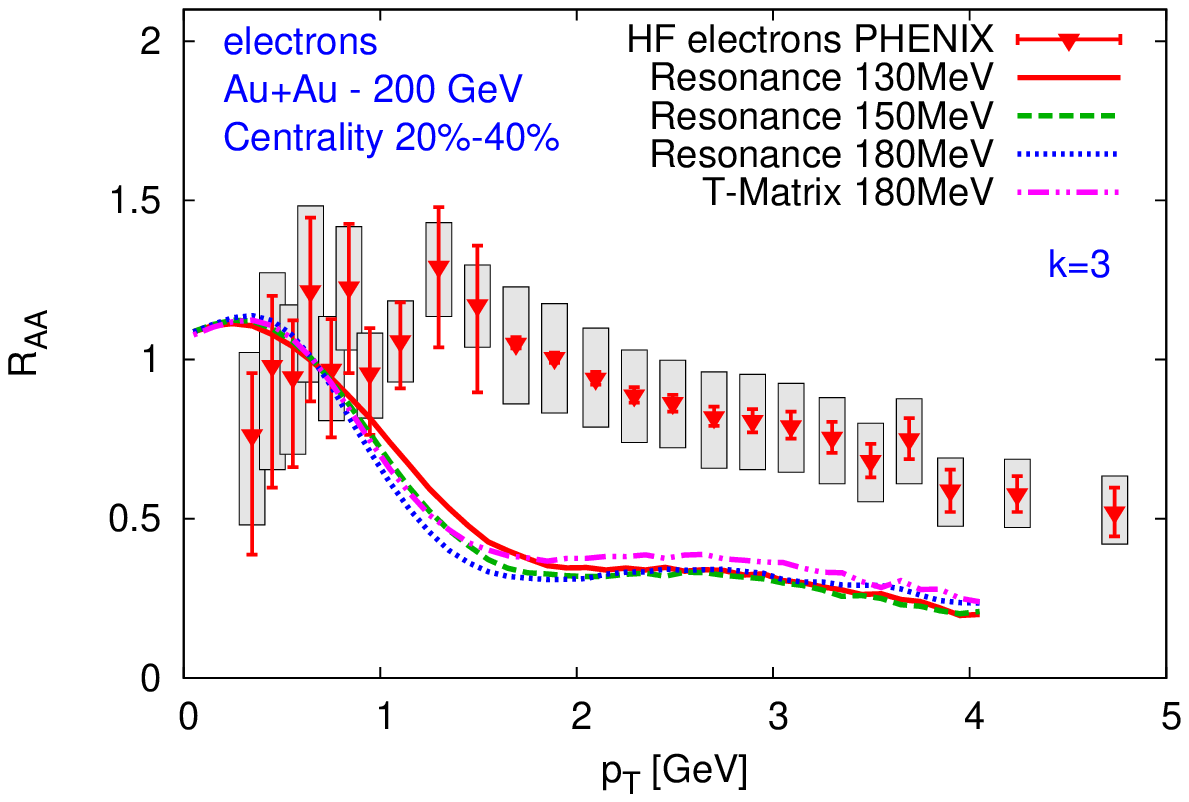}
\end{minipage}
\caption{(Color online) Left: $R_{AA}$ of electrons from heavy-flavor
  decays in Au+Au collisions at $\sqrt {s_{NN}}=200\;\GeV$ employing a
  $k$ factor of 3.  We use a rapidity cut of $|y|<0.35$. The solid lines
  show the results for charm quarks and the dashed ones for bottom
  quarks.\\
  Right: $v_2$ of electrons from heavy-flavor decays in Au+Au collisions
  at $\sqrt {s_{NN}}=200\;\GeV$ employing a $k$ factor of 3.  We use a
  rapidity cut of $|y|<0.35$. As expected the medium modification is
  stronger than without a $k$ factor. Data are taken from
  \cite{Adare:2010de}.  }
\label{k3RAARHIC4}
\end{figure}

Our results for the nuclear modification factor, $R_{AA}$, are depicted
in Fig.~\ref{k3RAARHIC4}.  The quenching is much stronger than
for the calculation without a $k$ factor.  Fig.~\ref{k3RAARHIC4} shows
the results for electrons.  The suppression of non-photonic electrons at
high $p_T$ is also stronger than for the calculation without a $k$
factor.

We conclude that the use of a $k$ factor can improve the description of
the elliptic flow.  However, it is not possible to reach a consistent simultaneous
description of both, the elliptic flow and the nuclear modification
factor using the same $k$ factor.

\subsection{Elliptic flow $v_2$ and nuclear modification factor $R_{AA}$ using Coalescence}

Instead of describing heavy quark hadronization by Peterson
fragmentation (and/or and additional k-factor, as discussed above) one
can alternatively apply a quark coalescence approach for D- and B-meson
production. To implement this coalescence we perform the Langevin
calculation until the decoupling temperature is reached. Subsequently we
coalesce a light quark with a heavy quark. As the light quarks
constitute the medium propagated by hydrodynamics, the average
velocities of the light quarks can be (on average) approximated by the 
flow-velocities of the hydro cells. 
The mass of the light quarks is assumed to be $370\;\MeV$ so that the
D-meson mass becomes $1.87\;\GeV$ when the masses of the light quarks
and the charm quarks ($1.5\;\GeV$) are added. Since we assume the light
quarks to have the same mass when coalescing with bottom quarks
($4.5\;\GeV$), the B-mesons obtain a mass of $4.87\;\GeV$.

\begin{figure}[h]
\begin{minipage}[b]{0.45\textwidth}
\includegraphics[width=1\textwidth]{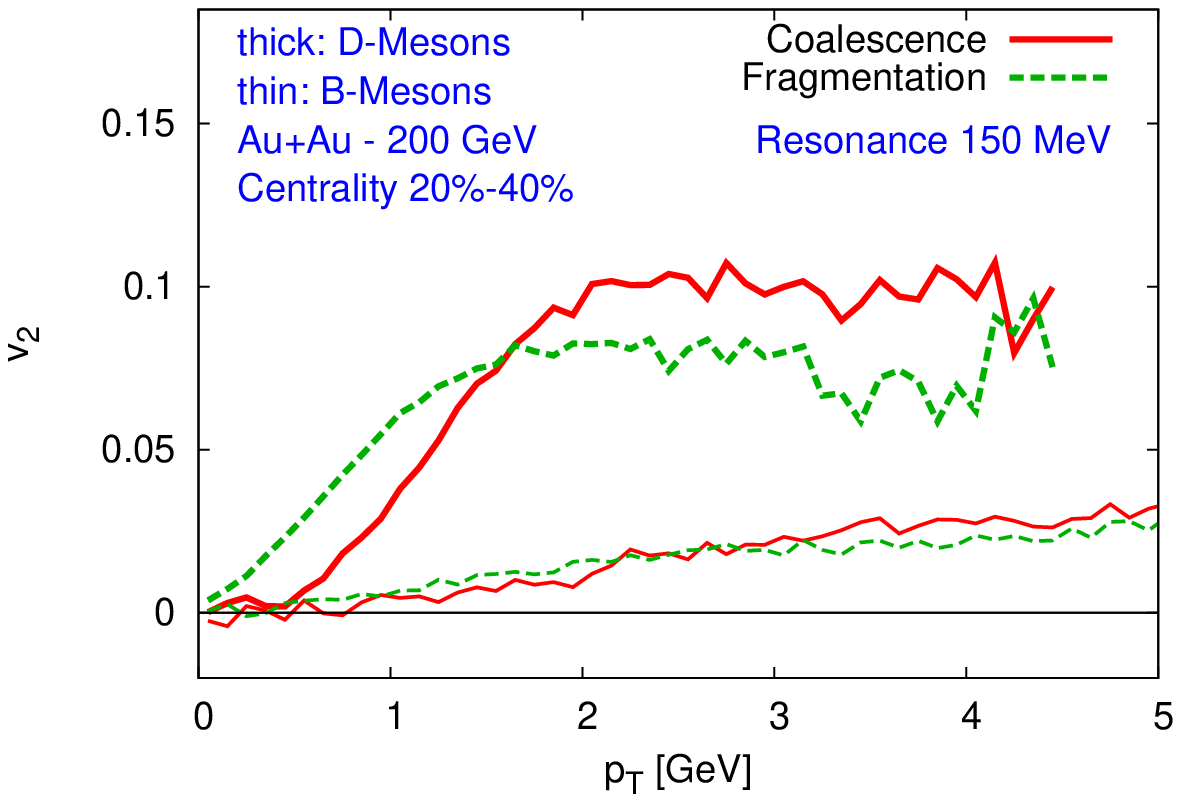}
\end{minipage}
\begin{minipage}[b]{0.45\textwidth}
\includegraphics[width=1\textwidth]{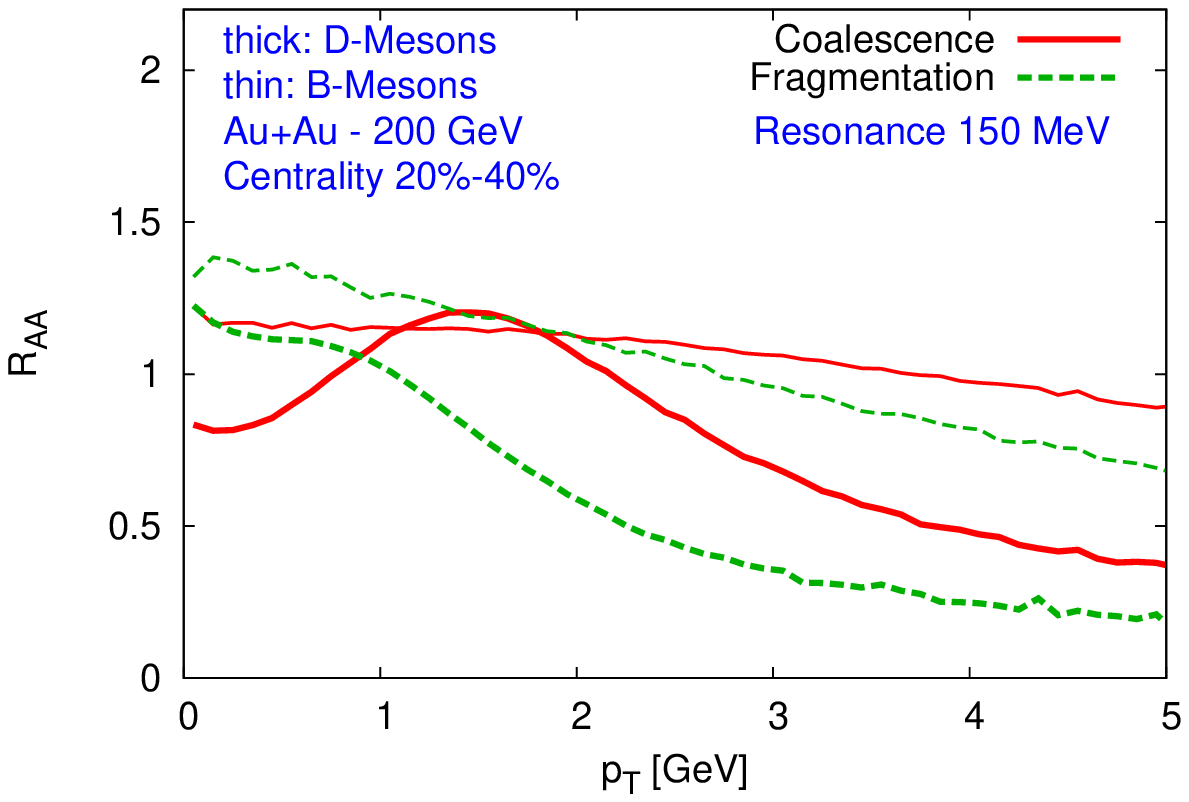}
\end{minipage}
\caption{(Color online) Elliptic flow $v_2$ (left) and $R_{AA}$ (right) of D mesons (solid lines)
  and B mesons (dashed lines) in Au+Au collisions at $\sqrt
  {s_{NN}}=200\;\GeV$.  We use a rapidity cut of $|y|<0.35$. A
  comparison of a Peterson fragmentation and a coalescence with light
  quarks is shown. For the drag and diffusion coefficients we use the resonance model with a decoupling temperature of $150\,\text{MeV}$. }
\label{CoaRAARHIC3}
\end{figure}

The differences of the flow and the spectra of D- and B-mesons when
comparing Peterson fragmentation (without $k$-factor) to the coalescence
model is shown in Fig. \ref{CoaRAARHIC3}.  These calculations are
performed employing a decoupling temperature of $150\,\text{MeV}$.

As compared to the fragmentation case, the elliptic flow reaches higher
values at high $p_T$ due to the coalescence. Also the depletion effect
described before is more pronounced.  Regarding the nuclear modification
factor, Fig. \ref{CoaRAARHIC3}, the difference of Peterson fragmentation
and the coalescence model is even larger. The push of low-$p_T$
particles to higher $p_T$ is stronger in case of the coalescence model,
while the suppression of heavy mesons at high $p_T$ is stronger in case
of Peterson fragmentation.

\begin{figure}[h]
\begin{minipage}[b]{0.45\textwidth}
\includegraphics[width=1\textwidth]{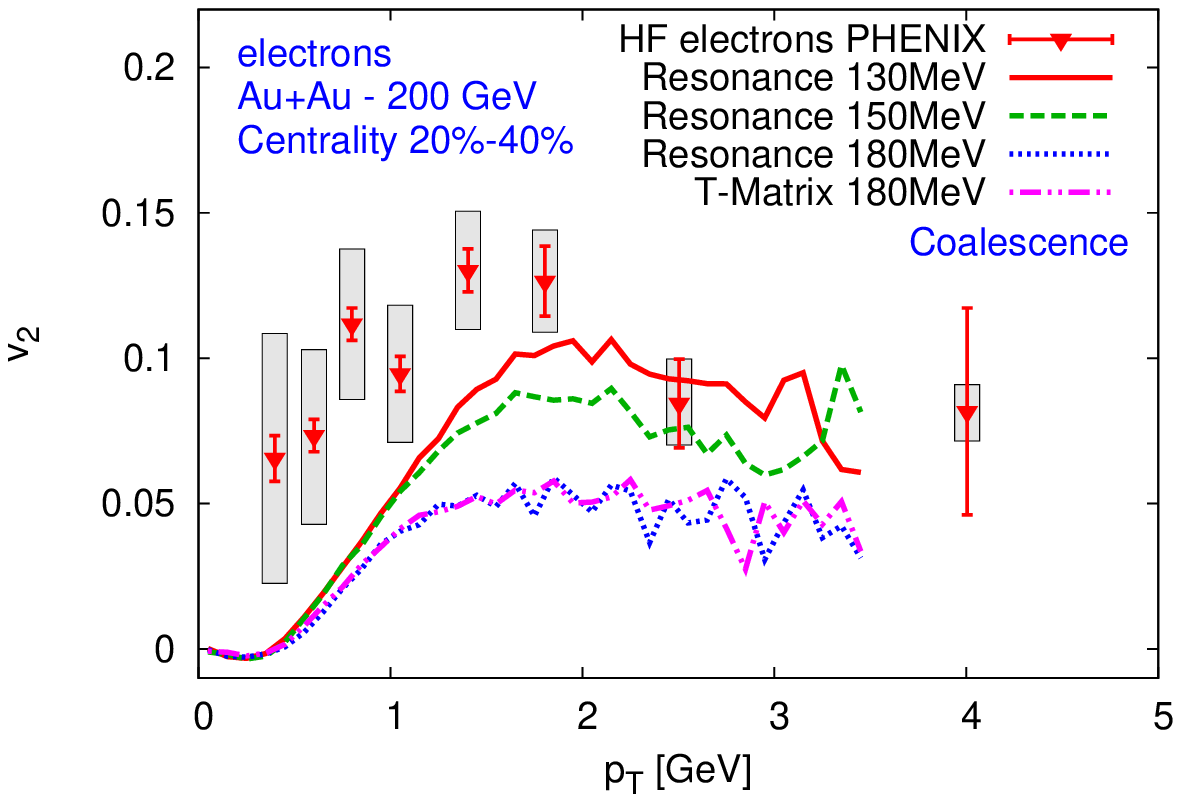}
\end{minipage}
\begin{minipage}[b]{0.45\textwidth}
\includegraphics[width=1\textwidth]{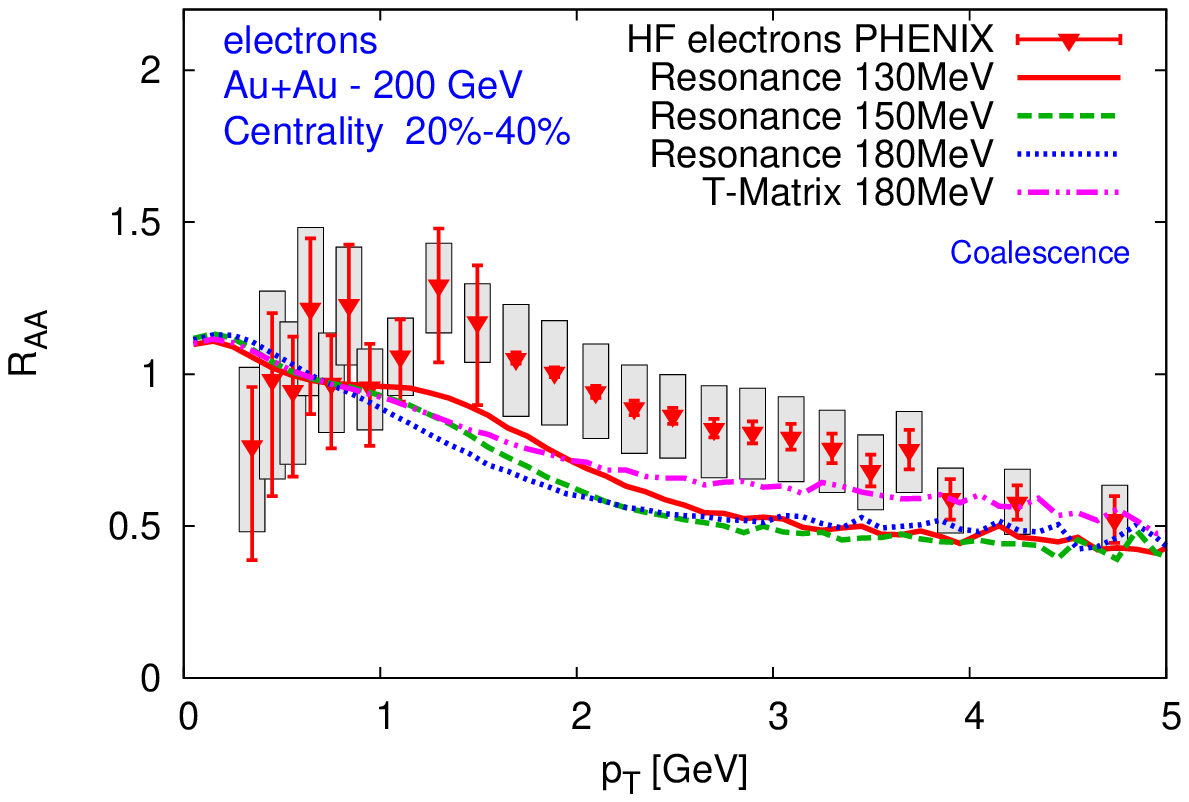}
\end{minipage}
\caption{(Color online) Elliptic flow $v_2$ (left) and nuclear modification factor 
$R_{AA}$ (right) of electrons from heavy quark decays
  in Au+Au collisions at $\sqrt {s_{NN}}=200\;\GeV$ using a coalescence
  mechanism.  We use a rapidity cut of $|y|<0.35$. For a decoupling
  temperature of $130\;\MeV$ we get a reasonable agreement to data
  \cite{Adare:2010de}.  }
\label{coaRAARHIC4}
\end{figure}

Again we perform a decay to electrons using PYTHIA to compare to
experimental measurements from the PHENIX
collaboration. Fig.~\ref{coaRAARHIC4} (left) shows our results for
$v_2$.  Due to the coalescence the elliptic flow is strongly increased
compared to the previous calculation using the Peterson
fragmentation. This higher flow is due to the momentum kick of the light
quarks in the recombination process, which provides additional flow from
the medium. For a decoupling temperature of $130\;\MeV$ we obtain a
reasonable agreement with the experimental data.

In Fig. \ref{coaRAARHIC4} (right) the nuclear modification factor for non-photonic 
single electrons is depicted.
Also here we obtain a good agreement with the data. Especially at
moderate $p_T\sim 2\,\text{GeV}$ the calculation has strongly improved.
The coalescence mechanism pushes the heavy quarks to higher $p_T$. As
seen before we obtain the best agreement to data for rather low
decoupling temperatures.

In conclusion we observe that the coalescence mechanism is required to
describe experimental data with our Langevin model.  Only with the
coalescence model one is able to describe both $R_{AA}$ and $v_2$
consistently in the present model.

\subsection{Dependence of the medium modification on the equation of state \label{secEOS}}

\begin{figure}[h]

\begin{minipage}[b]{0.45\textwidth}
\includegraphics[width=1\textwidth]{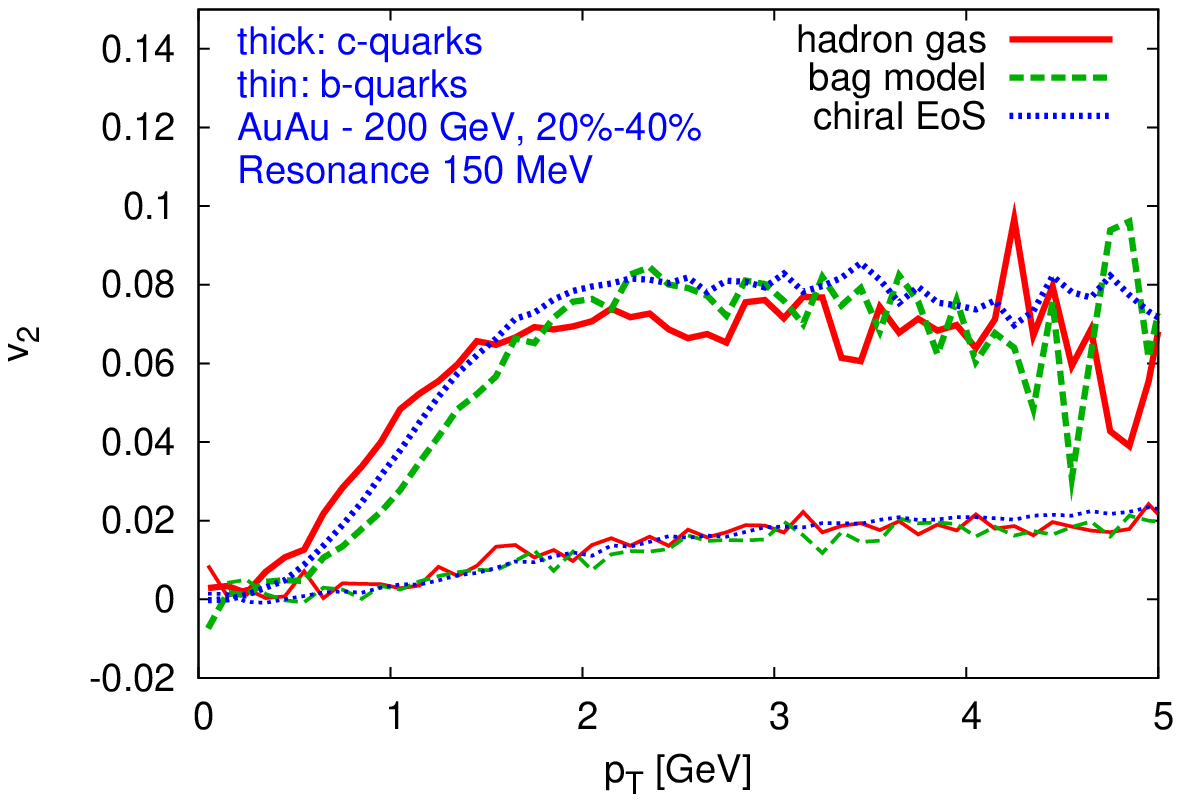}
\end{minipage}
\begin{minipage}[b]{0.45\textwidth}
\includegraphics[width=1\textwidth]{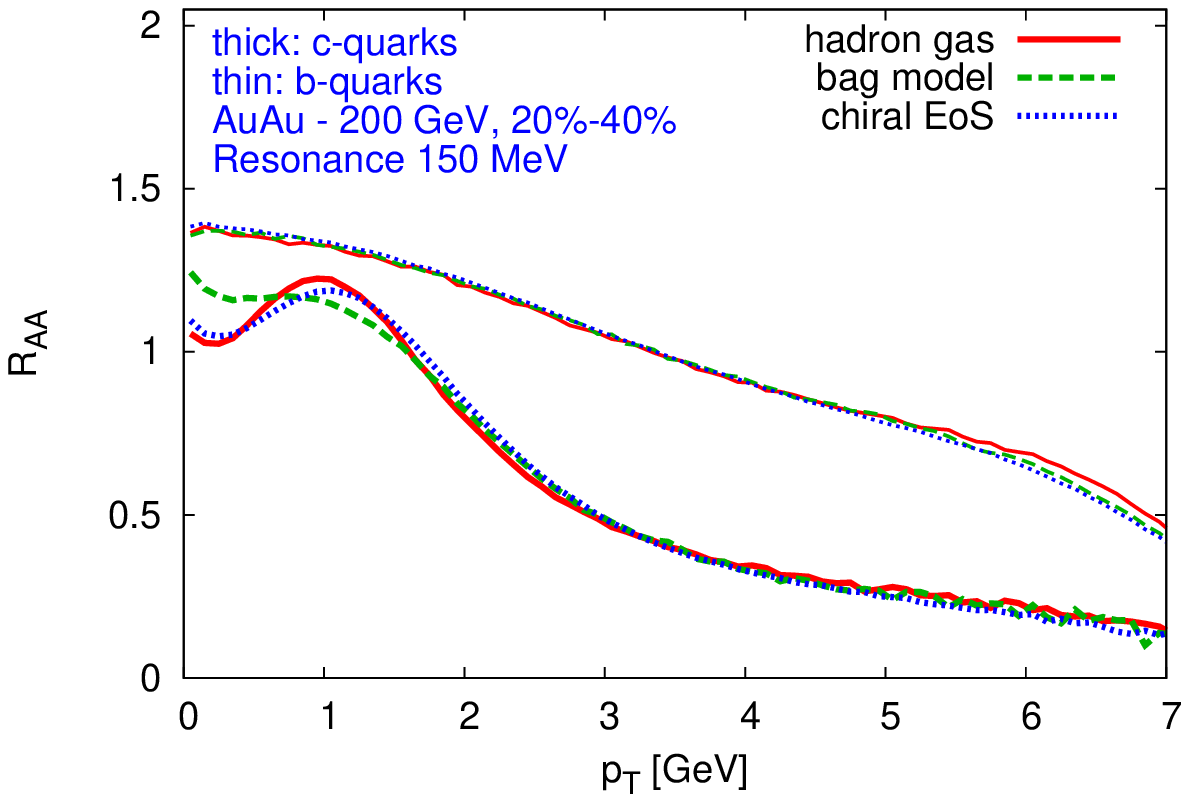}
\end{minipage}
\caption{(Color online) Elliptic flow $v_2$ (left) 
and nuclear modification factor $R_{AA}$ (right) of heavy quarks in Au+Au
  collisions at $\sqrt {s_{NN}}=200\,\GeV$. We use a rapidity cut of
  $|y|<0.35$. Different equations of state are compared.}
\label{RAAEOS}
\end{figure}

The heavy-flavor-flow observables in Langevin simulations are quite
sensitive to the used description of the background medium
\cite{Gossiaux:2011ea}. To examine this issue somewhat further, we have
performed our calculations also using different equations of state that
are implemented in the UrQMD hybrid model. Our results for different
equations of state for the drag and diffusion coefficients of the
resonance model with a decoupling temperature of $150\,\MeV$ are shown
in Fig.~\ref{RAAEOS} for the elliptic flow $v_2$ and for the nuclear
modification factor $R_{AA}$.

The equation of state we have been using for all results in the previous
sections is the chiral EoS. It is constructed by matching a state of the
art hadronic chiral model to a mean field description of the deconfined
phase. The deconfinement transition in this approach is included by the
means of an effective Polyakov Loop potential, coupling to the free
quarks. It has been shown in \cite{Steinheimer:2011ea} that the chiral
EoS gives a reasonable description of lattice QCD thermodynamics at
$\mu_B=0$ and can be extended to finite baryon densities. The
Hadron resonance gas EoS resembles the active degrees of freedom that
are also included in the UrQMD transport approach, namely most hadronic
states and their resonances. The Bag model EoS \cite{Rischke:1995mt}
follows from matching a Walecka type hadronic model to massless quarks
and gluons via a Maxwell construction. It exhibits a strong first order
phase transition for all values of $\mu_B$.

As one sees clearly the influence on the mediums evolution as seen through the heavy quarks for this set of
equations of state is very small.

\section{Results at LHC energies}

In the previous sections we found that we reach the best agreement to
experimental PHENIX data when using the Resonance model applying a
decoupling temperature of $130\,\text{MeV}$ and using quark coalescence
as hadronization mechanism. Now we apply the same description also at
LHC energies ($\sqrt{s_{NN}}=2.76\; \TeV$).  The momentum distribution
for the initially produced charm quarks at LHC is obtained from a fit to
PYTHIA calculations. The fit function we use is
\begin{equation}
\frac{\dd N}{\dd^2 p_T} =\frac{1}{(1+A_1\cdot \left(p_T^2\right)^{A_2})^{A_3}}
\end{equation}
with the coefficients $A_1=0.136$, $A_2=\,2.055$ and $A_3=\,2.862$.

We have performed our calculations in Pb+Pb collisions at $\sqrt{s_{NN}}
=2.76\; \TeV$ in a centrality range of 30\%-50\%. The analysis is done
in a rapidity cut of $|y|<0.35$ in line with the ALICE data.

\begin{figure}[h]
\begin{minipage}[b]{0.45\textwidth}
\includegraphics[width=1\textwidth]{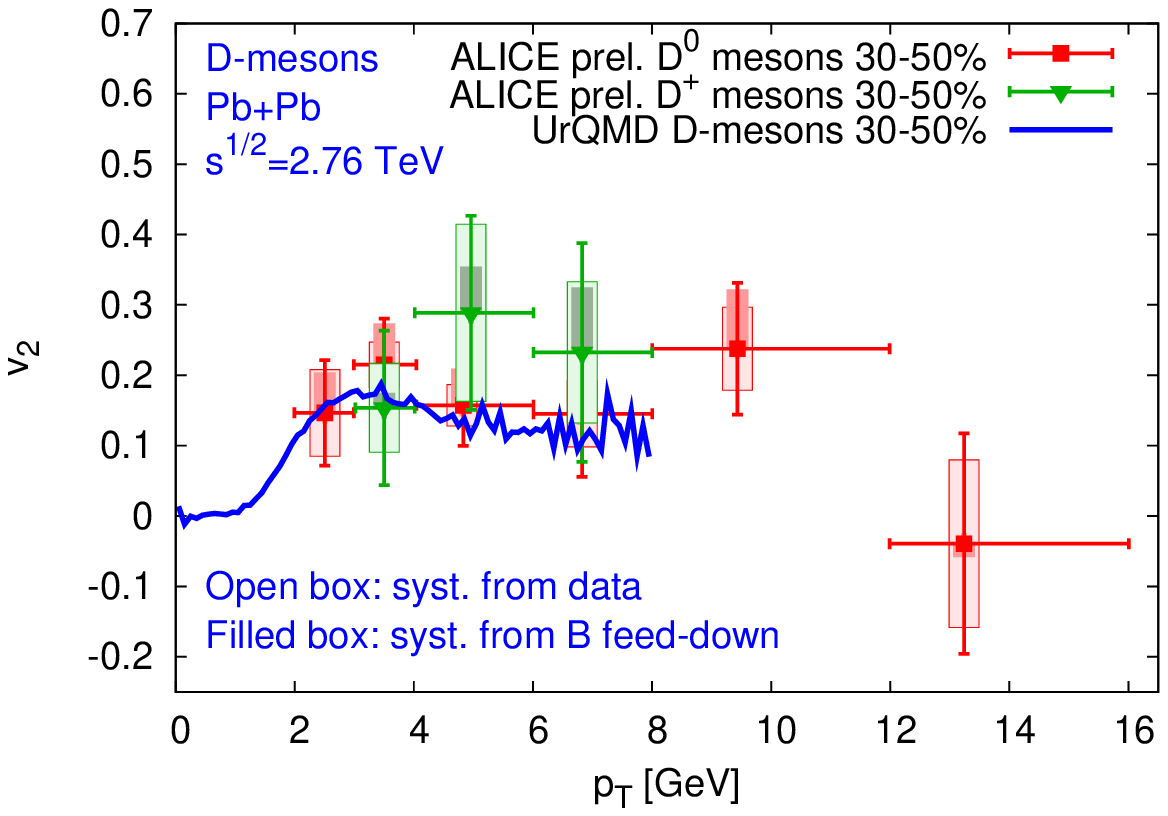}
\end{minipage}
\begin{minipage}[b]{0.45\textwidth}
\includegraphics[width=1\textwidth]{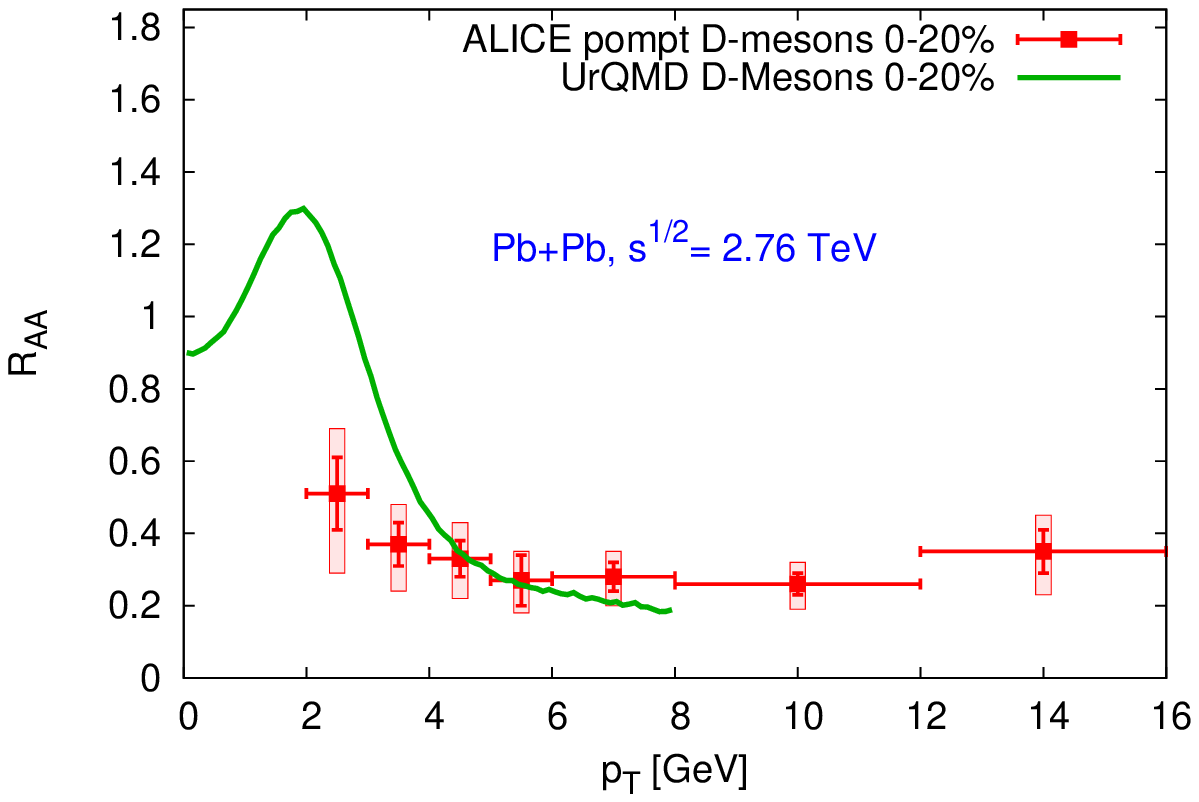}
\end{minipage}
\caption{(Color online) Left: Flow $v_2$ of D-mesons in Pb+Pb collisions at $\sqrt
  {s_{NN}}=2.76\,$TeV compared to data from the ALICE experiment. 
(Talk by Z.~Conesa del Valle at QM 2012, data not published yet.) 
A rapidity cut of $|y|<0.35$ is employed.\\ 
Right: $R_{AA}$ of D-mesons in Pb~Pb collisions at $\sqrt
  {s_{NN}}=2.76\; \TeV$ compared to experimental data from ALICE
  \cite{ALICE:2012ab}. A rapidity cut of $|y|<0.35$ is employed.}
\label{LHC4}
\end{figure}
Fig.\ \ref{LHC4} (left) depicts our results for the elliptic flow compared
to ALICE measurements. 
The D-meson $v_2$ exhibits a strong increase and reaches a maximum at
about $p_T=3\; \GeV$ with $v_2\sim 19\%$. The
agreement between the ALICE measurements of $D^0$- and $D^+$-mesons 
and our calculation is quite satisfactory.

A complementary view on the drag and diffusion coefficients is provided
by the nuclear suppression factor $R_{AA}$.  Figure \ref{LHC4} (right) shows
the calculated nuclear modification factor $R_{AA}$ of D-mesons at
LHC. In line with the experimental data the simulation is done for a
more central bin of $\sigma/\sigma_{\text{tot}}=0\%$-$20\%$. 
We find a maximum of the $R_{AA}$ at about $p_T=2\; \GeV$ followed by a
sharp decline to an $R_{AA}$ of about $0.2$ at high $p_T$. 
As we can see we can describe the
data at medium $p_T$ well but over-predict them at low $p_T$ bins.

\section{Summary}

In this paper we have investigated the medium modification of
heavy-quark $p_T$ spectra in the hot medium created in heavy-ion
collisions at RHIC and LHC energies on the basis of a Langevin
simulation coupled to the UrQMD hybrid model.  The aim of this study was
to find a consistent description for both the elliptic flow, $v_2$, and
the nuclear modification factor, $R_{AA}$, with a realistic dynamical
description of the background medium. We have used two different sets of
drag and diffusion coefficients, based on a $T$-Matrix approach and a
resonance-scattering model for the elastic scattering of heavy quarks
with light quarks and antiquarks. Both sets of coefficients lead to
similar results for the heavy-flavor observables.

In the first part of our analysis we have used a fragmentation
mechanism, the Peterson fragmentation, to describe the hadronization of
heavy quarks to open-heavy-flavor mesons. We have found a low elliptic
flow and a too strong heavy-flavor suppression at high $p_T$.
Subsequently we have explored how a $k$ factor for the drag and
diffusion coefficients would influence the results.  We found that with
$k=3$, the description of $v_2$ is improved, but has lead to an even
larger suppression of the nuclear modification factor $R_{AA}$, as
expected. We conclude that a combination of fragmentation and a Langevin
simulation with a $k$-factor in the transport coefficient does not allow
for a consistent description of the data on non-photonic single electron
spectra in Au+Au collisions ($\sqrt{s_{NN}}=200 \; \GeV$) at RHIC.

To overcome this problem we have used a coalescence approach to
heavy-quark hadronization to open-heavy-flavor mesons instead of the
fragmentation. The coalescence mechanism allows for a consistent
description of both $v_2$ and $R_{AA}$. We have performed the
simulations, assuming different decoupling temperatures of the heavy
quarks from the medium, and found that the late phase of the collision
can have a considerable effect on the heavy-quark observables. Within
our study we find the best agreement to experimental data using a low
decoupling temperature of $130\;\MeV$. In Sec.~\ref{secEOS} we have also
addressed the sensitivity of the heavy-flavor observables to the assumed
equation of state of the strongly interacting medium. Here we find that
our results are insensitive to variations of the particular equation of
state used in UrQMD's hydrodynamic model.

Finally we also explored the medium modification in our model at LHC
energies.  Here we could reach a good agreement to data for the elliptic
flow $v_2$ of D-mesons.  For the nuclear modification factor $R_{AA}$ we
reach a good agreement at medium $p_T$, but seem to miss the data at low
$p_T$ bins.

First measurements with the STAR Heavy Flavor Tracker (HFT) are
scheduled for the year 2014. It will provide new, complementary
measurements in heavy ion collisions at $\sqrt{s_{NN}}=200\,\text{GeV}$.
The HFT will enable direct identification of heavy flavor meson decays
like $D^0\rightarrow K^-\pi^+$ or $D_s^+\rightarrow K^-\pi^+K^+$. This
is supposed to lead to better $v_2$ measurements down to very low $p_T$
and a better understanding of the energy loss of heavy quarks in the
medium.  Especially it will provide us with identified D-meson spectra
which will enable us to compare our heavy-meson results to data
separately for D- and B-mesons and therefore to get further insights on
the hadronization mechanism.

\section*{ACKNOWLEDGMENTS}

T.\ Lang gratefully acknowledges support from the Helmholtz Research
School on Quark Matter Studies.  This work was supported by the Hessian
LOEWE initiative through the Helmholtz International Center for FAIR
(HIC for FAIR). J.\ S.\ acknowledges a Feodor Lynen fellowship of the
Alexander von Humboldt foundation. This work was supported by the Office
of Nuclear Physics in the US Department of Energy's Office of Science
under Contract No. DE-AC02-05CH11231 and the Bundesministerium f{\"ur}
Bildung und Forschung (BMBF) grant No. 06FY7083. The computational
resources were provided by the Frankfurt LOEWE Center for Scientific
Computing (LOEWE-CSC).

\newpage
\section{Appendix}
\subsection{Post-point Ito realization}
\label{AppA}

Since the phase-space distribution of relativistic particles is a scalar
\cite{deGroot80}, the proper equilibrium limit is given by the
corresponding boosted Boltzmann-J\"uttner phase-space distribution,
\begin{equation}
\label{lang.14}
f_Q^{(\text{eq})} \propto \exp \left (-\frac{p \cdot u}{T} \right),
\end{equation}
where $u(t,\bvec{x})$ is the four-velocity field of the medium and $p$
the (on-shell) four-momentum of the heavy quark in the local rest-frame.
It can be shown analytically, and we have numerically checked, that for
obeying this constraint, one has to apply the post-point prescription,
$\xi=1$, strictly only to the momentum argument of the covariance
matrix, $C_{jk}$ as given in (\ref{lang.4}) and not to the corresponding
coefficients originating from the Lorentz transformation of the time
step $\dd t$ with respect to the laboratory frame (bare symbols) to the
one in the local rest-frame of the heat bath (starred symbols), i.e., in
the transformation prescription for the time interval,
\begin{equation}
\label{lang.15}
\dd t^*=\frac{m}{E^*} \dd \tau = \frac{m}{E^*} \frac{E}{m} \dd t=\frac{E}{p \cdot u} \dd t,
\end{equation}
one has to use the heavy-quark momenta at time $t$ without a post-point
update rule. Here, $\dd \tau$ denotes the scalar ``proper-time''
interval of the heavy quark, corresponding to the given time interval,
$\dd t$, with respect to the laboratory frame \cite{Gossiaux:2011ea}.

\subsection{Drag and diffusion coefficients}
\label{AppB}

We use two non-perturbative models for elastic heavy-quark
scattering in the quark-gluon plasma to evaluate the drag and diffusion
coefficients for the Langevin simulation of heavy-quark diffusion.

The resonance model is based on heavy-quark effective theory (HQET) and
chiral symmetry in the light-quark sector
\cite{vanHees:2004gq}. Motivated by the finding in lattice-QCD
calculations that hadron-like bound states and/or resonances might
survive the phase transition in both the light-quark sector (e.g.,
$\rho$ mesons) and heavy quarkonia (e.g., $J/\psi$), in this model we
assume the existence of open-heavy-heavy-flavor meson resonances like
the D and B mesons.

In the $T$-Matrix approach static in-medium quark-antiquark potentials
from lattice QCD are used as scattering kernels in a Br\"uckner like
$T$-matrix approach to calculate the scattering-matrix elements for
elastic scattering of heavy quarks with light quarks and antiquarks
\cite{vanHees:2007me}.

The heavy-light quark resonance model\cite{vanHees:2004gq} is based on
the Lagrangian,
\begin{equation}
\begin{split}
\Lag_{Dcq} =& \Lag_D^0 + \Lag_{c,q}^0 - \ii G_S \left( \bar q \Phi_0^*
\frac{1+\fslash{v}}{2} c - \bar q \gamma^5 \Phi \frac{1+\fslash{v}}{2}
 c + h.c. \right)
\\
& - G_V \left( \bar q \gamma^{\mu} \Phi_{\mu}^* \frac{1+\fslash{v}}{2} c -
  \bar q \gamma^5 \gamma^{\mu} \Phi_{1\mu} \frac{1+\fslash{v}}{2} c + \text{h.c.}
\right) ,
\end{split}
\label{L_hq-eff}
\end{equation}
and an equivalent one for bottom quarks. Here $v$ denotes the
heavy-quark four-velocity. The free part of the Lagrangian is given by
\begin{equation}
\begin{split}
\Lag_{c,q}^0 &= \bar{c}(\ii \fslash{\partial}-m_c) c+\bar{q} \, \ii
\fslash{\partial} q,\\
\Lag_D^0 & =  (\partial_{\mu} \Phi^{\dagger})(\partial^{\mu} \Phi) +
(\partial_{\mu} {\Phi_0}^{*\dagger})(\partial^{\mu} \Phi_0^*)
-m_S^2(\Phi^{\dagger} \Phi+\Phi_0^{*\dagger} \Phi_0^*) \\
& \quad -\frac{1}{2} (\Phi_{\mu \nu}^{*\dagger} \Phi^{*\mu \nu}
+ \Phi_{1 \mu \nu}^{\dagger}
\Phi_1^{\mu \nu}) + m_V^2 (\Phi_{\mu}^{*\dagger} \Phi^{*\mu} +
\Phi_{1 \mu}^{\dagger} \Phi_1^{\mu}),
\end{split}
\label{L_0-eff}
\end{equation}
where $\Phi$ and $\Phi_0^*$ are pseudo-scalar and scalar meson fields
(corresponding to D and $\mathrm{D}_0^*$ mesons). Based on chiral
symmetry, restored in the QGP phase, we also assume the existence of
mass degenerate chiral-partner states. Further from heavy-quark
effective symmetry one expects spin independence for both the masses,
$m_S=m_V$, and the coupling constants, $G_S=G_V$. For the strange-quark
states we take into account only the pseudo-scalar and vector states
($D_s$ and $D_s^*$, respectively).

The D-meson propagators are dressed with the corresponding one-loop self
energy. Assuming charm- and bottom-quark masses of
$m_c=1.5\;\mathrm{GeV}$ and $m_b=4.5 \; \mathrm{GeV}$, we adjust the
masses of the physical D-meson-like resonances to $m_D=2 \; \GeV$ and
$m_B = 5 \; \GeV$, in approximate agreement with the $T$-matrix models
of heavy-light quark interactions in
\cite{Blaschke:2002ws,Blaschke:2003ji}. The coupling constant is chosen
such as to obtain resonance widths of $\Gamma_{D,B} =0.75 \; \GeV$.

With these propagators the elastic $Qq$- and $Q\overline{q}$-scattering
matrix elements are calculated and used for evaluation of the pertinent
drag and diffusion coefficients for the heavy quarks, using
(\ref{2.1.10b}) and (\ref{2.1.10c}). It turns out that particularly the
$s$-channel processes through a D/B-meson like resonance provide a large
efficiency for heavy-quark diffusion compared to the pQCD cross sections
for the same elastic scattering processes. This results in charm-quark
equilibration times $\tau_{\text{eq}}^c =2$-$10 \; \fm/c$.

In order to justify the formation of D- and B-meson like resonances
above $T_c$, in \cite{vanHees:2007me} a Brueckner-like in-medium
$T$-matrix approach has been used for the description of elastic
heavy-light-quark scattering in the QGP. After a three-dimensional
reduction to a Lippmann-Schwinger equation, including a Breit
correction, in-medium heavy-quark potentials from lQCD have been
employed as the scattering kernels. As an upper limit of the interaction
strength within such an approach, the internal-energy potential,
\begin{equation}
U(r,T)=F(r,T)-T \frac{\partial F(r,T)}{\partial T},
\end{equation} 
has been used, where $F$ is the free-energy potential from the lattice
calculation. We take into account also the complete set of $Q\bar{q}$
color states, assuming Casimir scaling of the corresponding potentials,
\begin{equation}
V_8=-\frac{1}{8} V_1, \quad V_{\overline{3}}=\frac{1}{2} V_1, \quad
V_6=-\frac{1}{4} V_1.
\end{equation}
After a partial-wave decomposition the Lippmann-Schwinger equation,
\begin{equation}
\begin{split}
\label{tmat}
T_{a,l}(E;q'& ,q)= V_{a,l}(q',q) \\
&+ \frac{2}{\pi} \int \dd k \; k^2
V_{a,l}(q',k) G_{Qq}(E,k) \\ 
& \times T_{a,l}(E;k,q) [1-f_F(\omega_k^Q) - f_F(\omega_k^q)],
\end{split}
\end{equation}
for the partial-wave components of each color channel, $a$, has been
solved for the $S$- and $P$-wave components. Here, $E$ is the
center-of-momentum (CM) energy of the heavy-light quark system, $q$ and
$q'$ the momenta of the heavy and light quark, and
\begin{equation}
G_{qQ}(E,k)=\frac{1}{E-(\omega_k^q+\ii \Sigma_I^q)-(\omega_k^Q+\ii
  \Sigma_I^Q)}
\end{equation}
the corresponding two-particle propagator in the CM frame. It has been
checked that the quasi-particle widths of $\Gamma_I^{q,Q}=2
\Sigma_I^{q,Q}=200 \; \MeV$ are consistent with a previous similar
Br"uckner calculation \cite{Mannarelli:2005pz} for the light quarks and
with the heavy-quark self-energies with the $T$-matrix solution of
(\ref{tmat}). The relation with the invariant scattering-matrix elements
in (\ref{2.1.10b}) is then given by
\begin{equation}
\begin{split}
\sum & |\mathcal{M}|^2 = \frac{64 \pi}{s^2}
(s-m_q^2+m_Q^2)^2(s-m_Q^2-m_q^2)^2 \\
& \times N_f \sum_{a} d_a \left (|T_{a,l=0}(s)|^2+ 3|T_{a,k=1}(s) \cos
  \theta_{\text{cm}}|^2 \right). 
\end{split}
\end{equation}
The relation of elastic heavy-quark-scattering matrix elements with the
drag and diffusion coefficients in the Langevin approach is given by
integrals of the form
\begin{equation}
\begin{split}
\label{2.1.10b}
\erw{X(\bvec{p}')}= &\frac{1}{2 \omega_{\bvec{p}}} \tildeint{\bvec{q}}
\tildeint{\bvec{p}'} \tildeint{\bvec{q}'} \frac{1}{\gamma_Q} \sum_{g,q} 
|\mathcal{M}|^2 \\
& \times (2 \pi)^4 \delta^{(4)}(p+q-p'-q') f_{q,g}(\bvec{q}) X(\bvec{p}') \ .
\end{split}
\end{equation}
Here, the integrations run over the three momenta of the incoming light
quark or gluon and the momenta of the outgoing particles. The sum over
the matrix element is taken over the spin and color degrees of freedom
of both the incoming and outgoing particles; $\gamma_Q=6$ is the
corresponding spin-color degeneracy factor for the incoming heavy quark,
and $f_{q,g}$ stands for the Boltzmann distribution function for the
incoming light quark or gluon. In this notation, the drag and diffusion
coefficients are given by
\begin{equation}
\begin{split}
\label{2.1.10c}
A(\bvec{p}) &= \erw{1-\frac{\bvec{p} \bvec{p}'}{\bvec{p}^2}}, \\
B_{0}(\bvec{p}) &= \frac{1}{4} \erw{\bvec{p}'{}^2-\frac{(\bvec{p}'
    \bvec{p})^2}{\bvec{p}^2}} , \\
B_{1}(\bvec{p}) &= \frac{1}{2} \erw{\frac{(\bvec{p'}
    \bvec{p})^2}{\bvec{p}^2} - 2 \bvec{p}' \bvec{p} + \bvec{p}^2}. 
\end{split}
\end{equation}

For both approaches we also include the leading-order perturbative QCD
cross sections for elastic gluon heavy-quark scattering
\cite{Combridge:1978kx}, including a Debye screening mass $m_{Dg}=g T$
in the gluon propagators, thus taming the $t$-channel singularities in
the matrix elements. The strong-coupling constant is chosen as
$\alpha_s=g^2/(4 \pi)=0.4$.

\newpage
\subsection{Underlying D- and B-meson spectra before semi-leptonic decays}
\label{AppC}

\begin{figure}[h]
\begin{minipage}[h]{0.45\textwidth}
\includegraphics[width=1\textwidth]{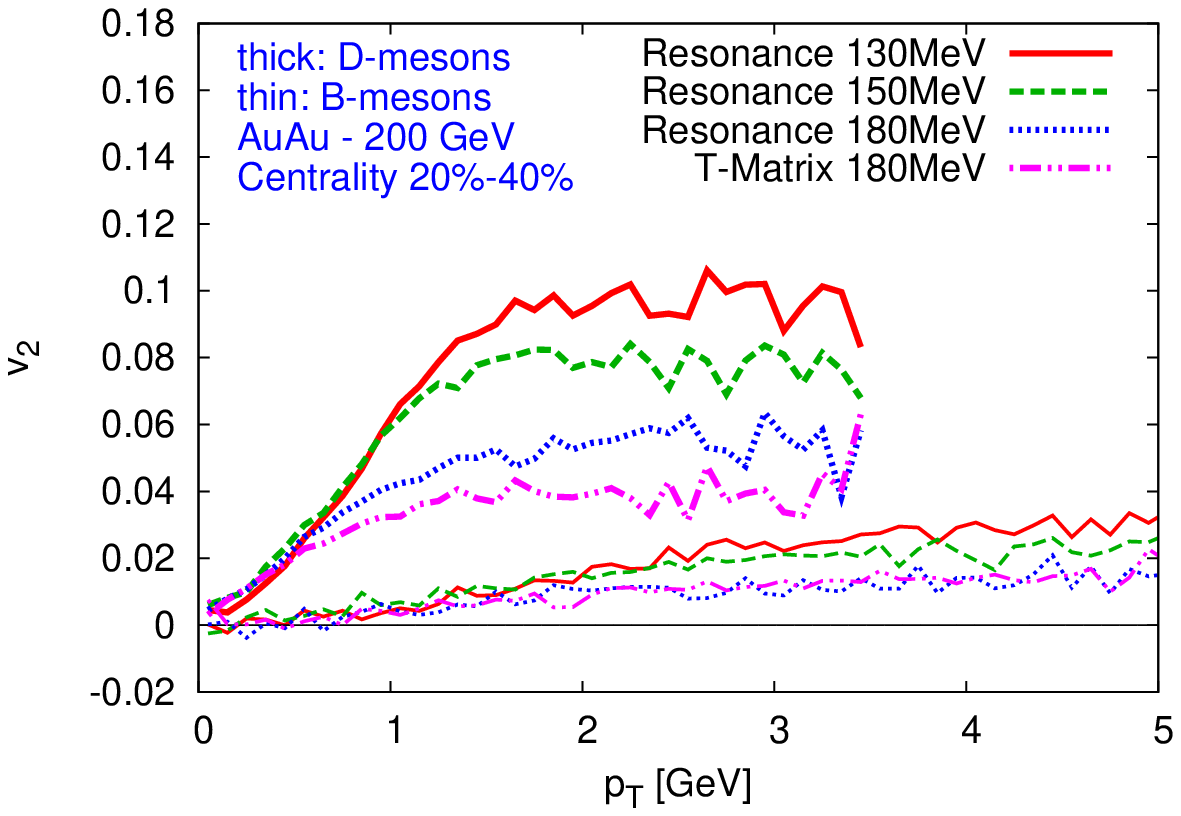}
\end{minipage}
\begin{minipage}[h]{0.45\textwidth}
\includegraphics[width=1\textwidth]{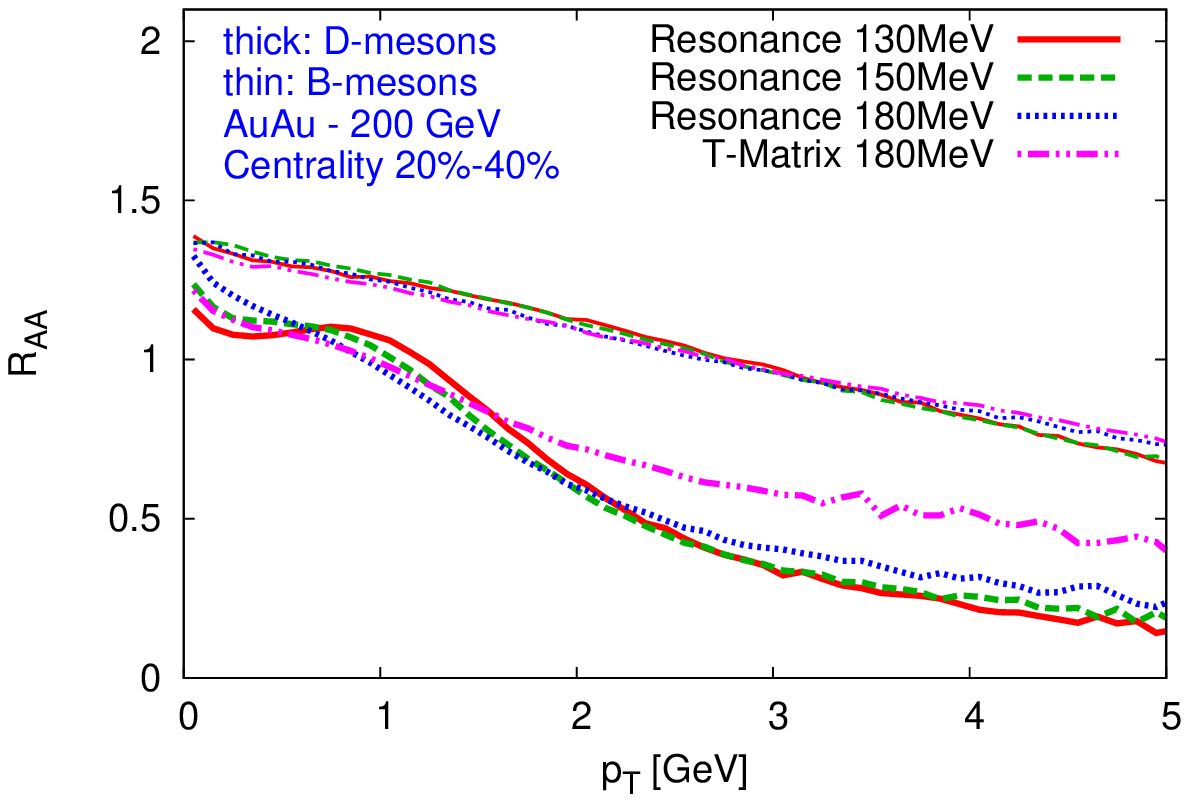}
\end{minipage}
\caption{(Color online) Elliptic flow $v_2$ (left) and nuclear
  modification factor $R_{AA}$ (right) of D- and B-mesons using Peterson
  fragmentation in Au+Au collisions at $\sqrt {s_{NN}}=200\,\GeV$. We
  use a rapidity cut of $|y|<0.35$.}
\label{MesonsNormal}
\end{figure}

\begin{figure}[h]
\begin{minipage}[h]{0.45\textwidth}
\includegraphics[width=1\textwidth]{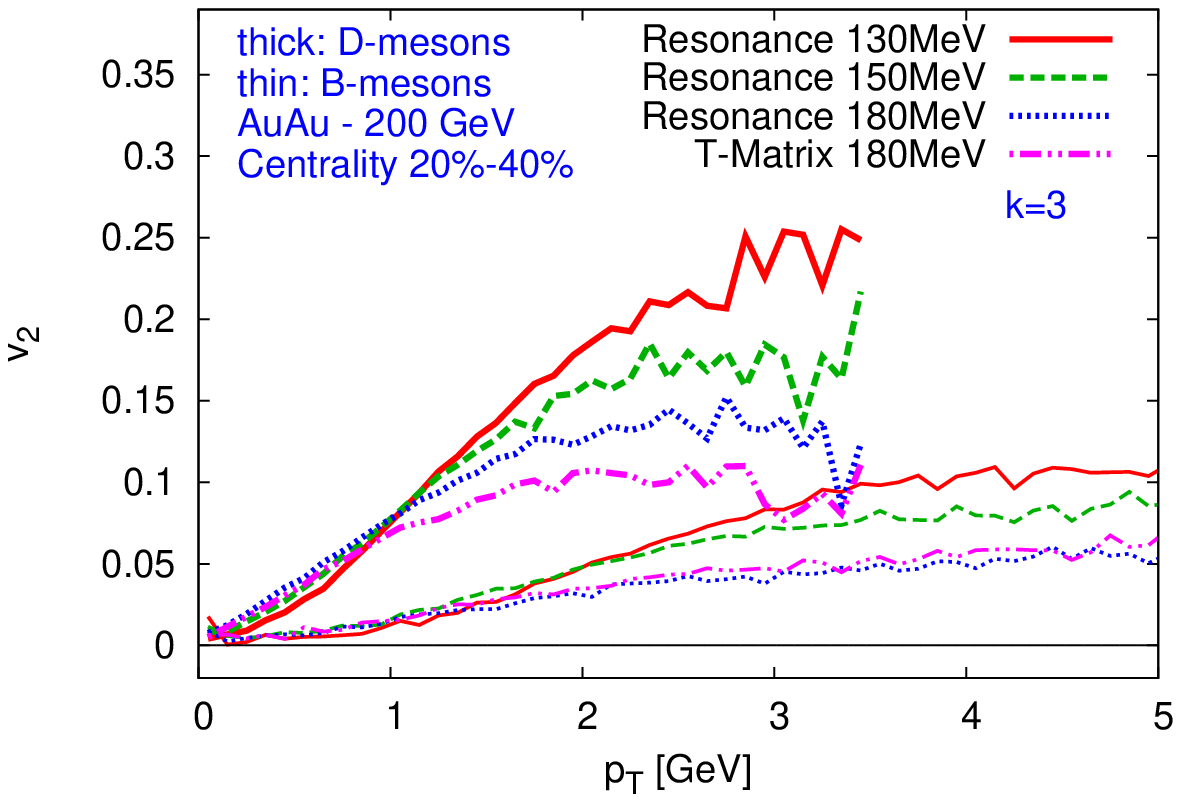}
\end{minipage}
\begin{minipage}[h]{0.45\textwidth}
\includegraphics[width=1\textwidth]{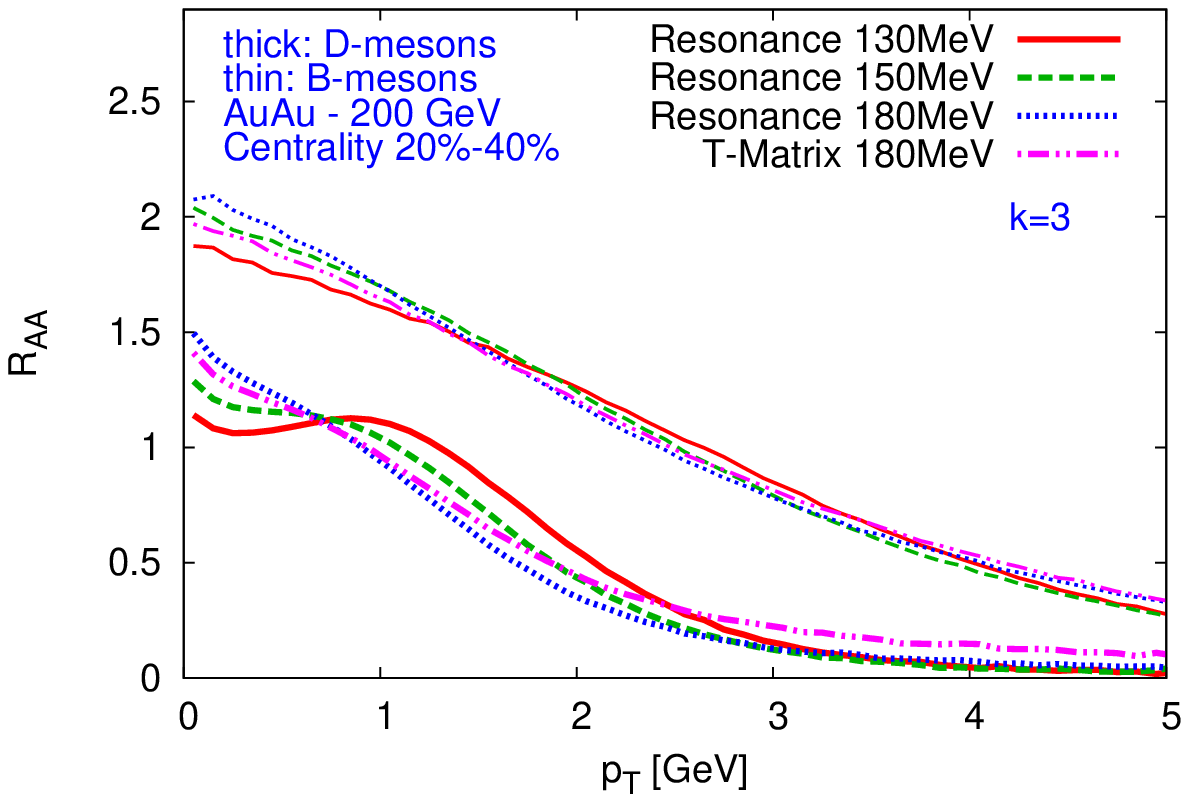}
\end{minipage}
\caption{(Color online) Elliptic flow $v_2$ (left) 
and nuclear modification factor $R_{AA}$ (right) of D- and B-mesons using Peterson fragmentation and a k-factor of 3 in Au+Au
  collisions at $\sqrt {s_{NN}}=200\,\GeV$. We use a rapidity cut of
  $|y|<0.35$.}
\label{Mesonsk3}
\end{figure}

\begin{figure}[h]
\begin{minipage}[h]{0.45\textwidth}
\includegraphics[width=1\textwidth]{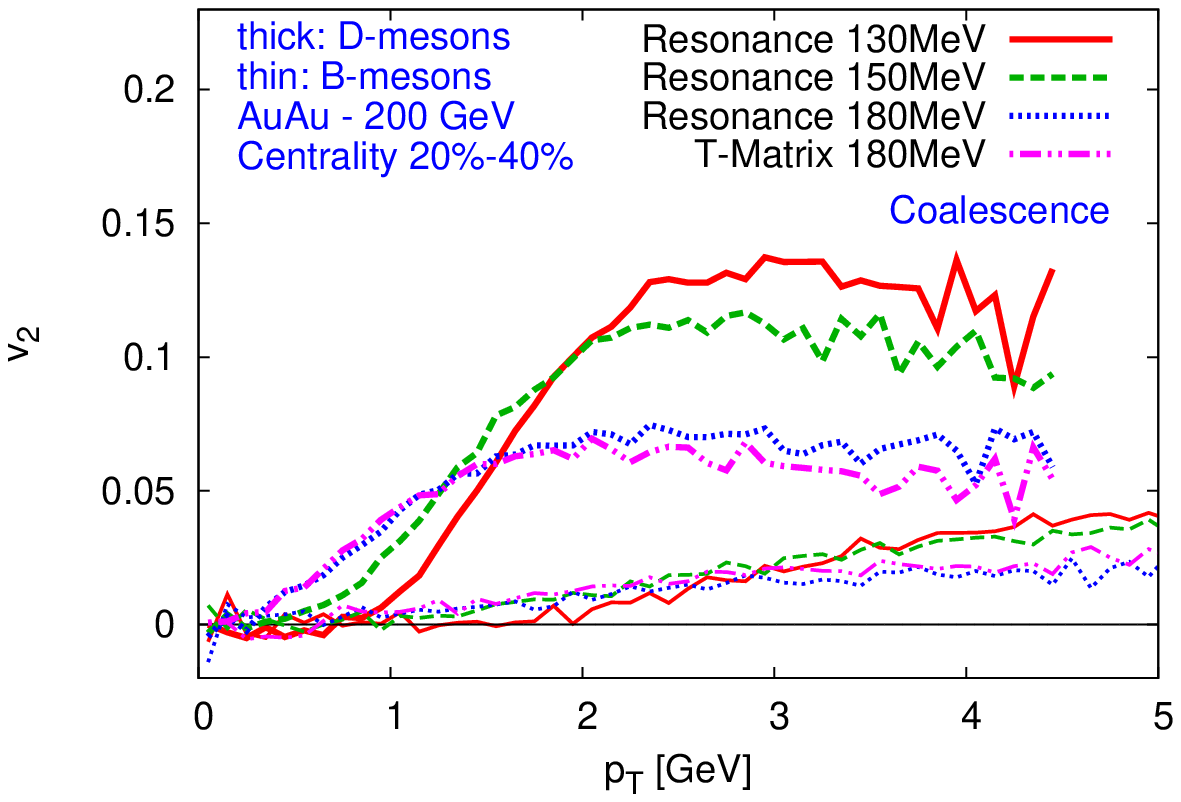}
\end{minipage}
\begin{minipage}[h]{0.45\textwidth}
\includegraphics[width=1\textwidth]{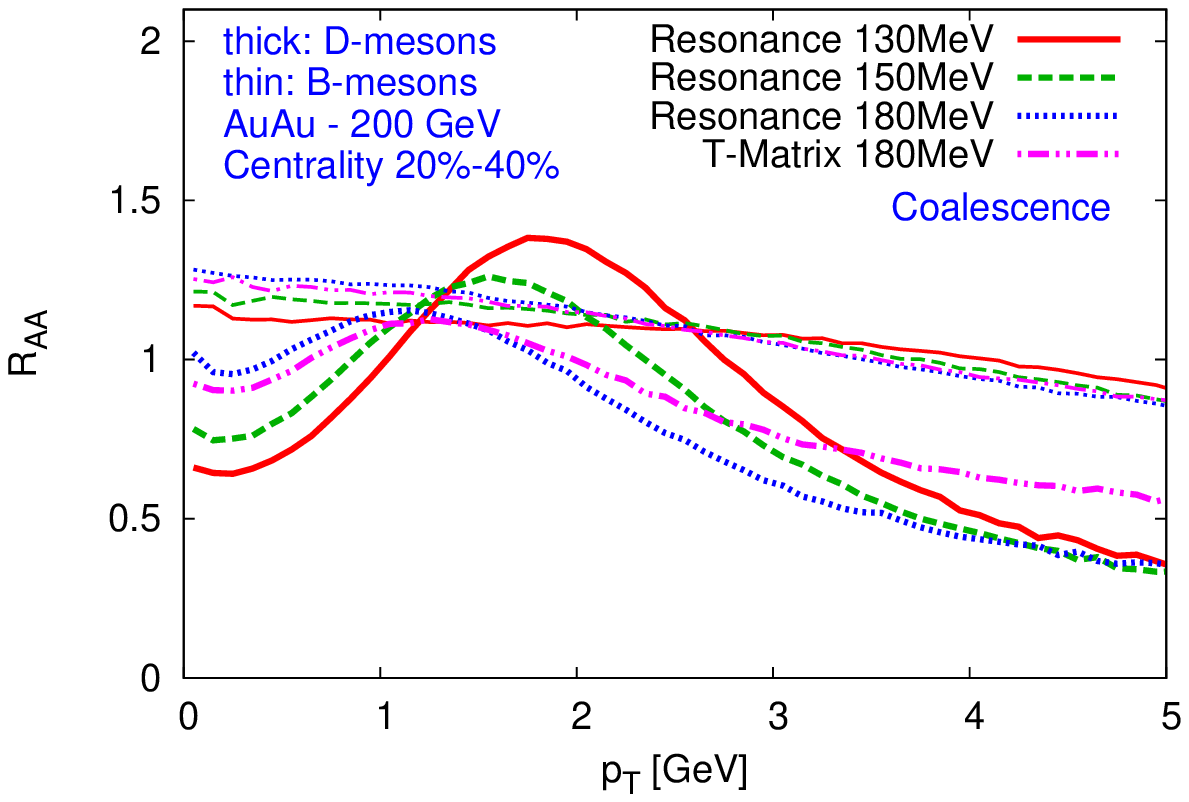}
\end{minipage}
\caption{(Color online) Elliptic flow $v_2$ (left) 
and nuclear modification factor $R_{AA}$ (right) of D- and B-mesons using coalescence in Au+Au
  collisions at $\sqrt {s_{NN}}=200\,\GeV$. We use a rapidity cut of
  $|y|<0.35$.}
\label{MesonsCoa}
\end{figure}

The heavy flavor electron spectra at RHIC originate from D- and B-meson
decays.  These D- and B-meson spectra are obtained from our heavy quark
calculations applying a fragmentation or a coalescence mechanism.  They
are displayed in Fig. \ref{MesonsNormal} for the case of the Peterson
fragmentation without using a $k$-factor, in Fig. \ref{Mesonsk3} for the
case of the Peterson fragmentation applying a $k$-factor of 3 and
finally for the case of using a coalescence mechanism
(Fig. \ref{MesonsCoa}).
 
These spectra can act as a prediction for future D- and B-meson measurements at RHIC energies. 
On the one hand they allow for a comparison of our hadronization mechanisms to experimantal data 
and on the other hand for a comparison of the decay to heavy flavor electrons performed using PYTHIA.


\newpage
\bibliography{bibliography}
\end{document}